\pdfoutput=1
\RequirePackage{ifpdf}
\ifpdf 
\documentclass[pdftex]{sigma}
\else
\documentclass{sigma}
\fi

\numberwithin{equation}{section}

\begin{document}

\newcommand{\lP}{\ell_{\mathrm P}}

\allowdisplaybreaks

\renewcommand{\PaperNumber}{082}

\FirstPageHeading

\ShortArticleName{Mathematical Structure of Loop Quantum Cosmology: Homogeneous Models}

\ArticleName{Mathematical Structure of Loop Quantum\\
Cosmology: Homogeneous Models}

\Author{Martin BOJOWALD}

\AuthorNameForHeading{M.~Bojowald}

\Address{Institute for Gravitation and the Cosmos, The Pennsylvania State University,\\
104 Davey Lab, University Park, PA 16802, USA}
\Email{\href{mailto:bojowald@gravity.psu.edu}{bojowald@gravity.psu.edu}}

\ArticleDates{Received August 08, 2013, in f\/inal form December 22, 2013; Published online December 30, 2013}

\Abstract{The mathematical structure of homogeneous loop quantum cosmology is analy\-zed, starting with and
taking into account the general classif\/ication of homogeneous connections not restricted to be Abelian.
As a~f\/irst consequence, it is seen that the usual approach of quantizing Abelian models using spaces of
functions on the Bohr compactif\/ication of the real line does not capture all properties of homogeneous
connections.
A new, more general quantization is introduced which applies to non-Abelian models and, in the Abelian
case, can be mapped by an isometric, but not unitary, algebra morphism onto common representations making
use of the Bohr compactif\/ication.
Physically, the Bohr compactif\/ication of spaces of Abelian connections leads to a~degeneracy of edge
lengths and representations of holonomies.
Lifting this degeneracy, the new quantization gives rise to several dynamical properties, including lattice
ref\/inement seen as a~direct consequence of state-dependent regularizations of the Hamiltonian constraint
of loop quantum gravity.
The representation of basic operators~-- holonomies and f\/luxes~-- can be derived from the full theory
specialized to lattices.
With the new methods of this article, loop quantum cosmology comes closer to the full theory and is in
a~better position to produce reliable predictions when all quantum ef\/fects of the theory are taken into
account.}


\Classification{81R10; 39A14}

\section{Introduction}

Loop quantum cosmology~\cite{LivRev,Springer} aims to develop and analyze cosmological models by
incorporating crucial guidance from the full theory of loop quantum gravity~\cite{ALRev, Rov,ThomasRev}.
Even though its systems cannot yet be derived completely, they constitute more than a~set of minisuperspace
models.
Characteristic ef\/fects of quantum geometry have been found in this setting, and contact with a~potential
full framework of quantum gravity has allowed one to f\/ix some choices left open in traditional models of
quantum cosmology.
Nevertheless, ambiguities remain in loop quantum cosmology and loop quantum gravity, to be described by
suf\/f\/iciently general parameterizations that might be restricted by phenomenological analysis.
Also for these parameterizations, contact with the full theory is essential: many dif\/ferent features
collapse on one single parameter when geometry is restricted to exact homogeneity.
Disentangling dif\/ferent contributors to one ef\/fect is important for estimates of typical ranges of
parameters.

Ambiguities notwithstanding, some general features have been found, foremost among them quantum
hyperbolicity~\cite{Sing,BSCG}, a~mechanism of singularity avoidance based on discrete structures of
evolution operators.
However, ef\/fective geometrical pictures of resolution mechanisms are dif\/f\/icult to derive since
several dif\/ferent quantum ef\/fects contribute, in addition to higher-curvature corrections also
quantum-geometry modif\/ications.
(Space-time as we know it may not even exist in extreme quantum phases~\cite{Action, ModifiedHorizon}.)
Specif\/ic details of physics in deep quantum regimes, for instance the values of upper bounds on the
energy density of matter, as another possible indicator of singularity resolution, are often unreliable
because the setting remains too reduced and too ambiguous.
An appropriate viewpoint is one akin to ef\/fective f\/ield theories, where one uses proposals for full
(but possibly incomplete) theories to derive generic low-energy properties.
With such a~view, the framework is empirically testable because it can give rise to potentially observable
phenomena, even if they cannot be predicted with certainty in all their parameterized details.

\looseness=-1 Developments in loop quantum cosmology have not always followed a~general view, especially
since the publication of~\cite{APS}, which stimulated a~line of research focusing with minutest detail on
pure minisuperspace models.
Strenuous contact with the full theory has been replaced by ad-hoc assumptions (for instance related to
degrees of freedom and scaling behaviors\footnote{See Sections~\ref{s:adhoc} and~\ref{s:Param}.} in
discrete structures); ambiguities (such as the so-called area gap and its postulated dynamical role) have
been f\/ixed by hand.
Valuable results have been produced, chief\/ly of mathematical interest, showing what discrete features and
non-standard quantum representations may imply; see e.g.\ \cite{SelfAdFlat,NonSelfAd,PhysEvolBI}.
For physical statements, however, the viewpoint espoused likely contains too many artefacts to be reliable.
This article adds additional items to the list of known minisuperspace limitations.

The main aim, however, is to provide a~general description of homogeneous reduced systems for quantum
cosmology, focusing on but not restricted to loop quantum gravity.
Since quantum cosmological models are beginning to be developed in approaches closely related to loop
quantum gravity, such as spin foams~\cite{SpinFoamRev, SurfaceSum,Rov:Loops}, it is important to state the
general setting of quantum cosmological models, and to point out limitations, dangers, and promises.
The following section begins with a~description and classif\/ication of symmetric models within full
(classical or quantum) theories, amended in later sections by specif\/ic details and discussions within
loop quantum cosmology and, brief\/ly, the spin-foam setting.
The f\/inal section will put these models in the general framework of ef\/fective theory.
Along the road, we will be led to several mathematical features overlooked so far.
Most importantly, Hilbert space representations based on functions on the Bohr compactif\/ication of the
real line do not properly capture all aspects of homogeneous connections; rather, they give rise to
a~degeneracy of two important parameters corresponding, in the full theory, to the edge length and
representation label of holonomies.
The origin of the degeneracy is identif\/ied here as a~mathematical coincidence realized in Abelian models
only.
To solve these problems, we will present a~new non-Abelian construction, work out a~detailed relation to
the full theory, and arrive at a~new viewpoint on dynamical dif\/ference equations with a~natural
implementation of lattice ref\/inement.

\section{Reduction}
\label{s:Reduction}

A classical reduced model, realizing a~given symmetry, uses an embedding ${\cal M}\to{\cal S}$ from the set~${\cal M}$ of symmetric geometries into full superspace~${\cal S}$.
A minisuperspace geometry as an element of ${\cal M}$ is specif\/ied by f\/initely many parameters~$a_I$
(which, to be specif\/ic, one may think of as the three scale factors of a~diagonal Bianchi model), mapped
to a~full metric $g_{ab}(a_I)$ by the classif\/ication of invariant metric tensors.
(A classif\/ication of invariant connections and triads is mathematically more
well-developed~\cite{Brodbeck, KobNom}; see~\cite{CUP, SymmRed} for applications in the present context.)
Inserting $g_{ab}(a_I)$ in equations of motion, the action or constraints of the full theory then produces
corresponding equations for the f\/initely many $a_I$.\footnote{The procedure of inserting symmetric
tensors may not commute with variations used to compute equations of motion~\cite{midisup2, midisup}:
symmetric actions or constraints do not always produce the correct symmetric equations of motion.
In our following discussions we make use of constraints, and therefore must assume that variation and
symmetrization commute.
In the Bianchi classif\/ication of homogeneous models, for instance, we are restricted to class
A~\cite{classAB}.}

We distinguish between a~minisuperspace model (a def\/inition of minisuperspace ${\cal M}$ with a~dynamical
f\/low on it) and the stronger notion of a~reduced model (a minisuperspace model including also an
embedding ${\cal M}\to{\cal S}$, making contact with the full theory).
There is not much of a~dif\/ference classically, but there is a~big one at the quantum level.
In quantum cosmology, most models remain in the minisuperspace setting, def\/ining some dynamical f\/low on
a~system with f\/initely many degrees of freedom such that the dynamics of general relativity follows in
a~semiclassical limit, perhaps with inspiration by but no derivation from some full theory that quantizes
${\cal S}$.
The key ingredient of reductions~-- making contact with a~full theory of quantum gravity, if only in
a~weak sense~-- most often is missing.
This article deals with the problem of reduction at the quantum level, going beyond pure minisuperspace
models.

\subsection{Reduction, selection, projection}

In addition to minisuperspace versus reduced models, it will be useful to distinguish between three
classical procedures of deriving symmetric systems beyond a~mere minisuperspace prescription:
\begin{description}\itemsep=0pt
\item[Reduction:] The def\/inition of a~reduced model, as already stated, contains two parts:
embedding ${\cal M}$ in ${\cal S}$ and deriving a~dynamical f\/low by inserting minisuperspace metrics into
the full equations.
After a~reduced model has been def\/ined, one can proceed to solve its equations and evaluate solutions for
potential predictions\footnote{``Embedding'' might be a~better name for this prescription, but
``reduction'' is much more standard.}.
\item[Selection:] Instead of starting with a~set of minisuperspace geometries and deriving a~reduced f\/low
from the full theory, one could f\/irst solve the full equations and then select symmetric ones that admit
a~set of Killing vector f\/ields with an algebra of the desired symmetry type.
In general relativity with its complicated and largely unknown solution space, this procedure is rather
impractical.
\item[Projection:] Or, again starting with solutions to the full equations, one may derive a~symmetric
geometry for every full solution by some kind of averaging process.
In addition to the problem of a~selection procedure, one would have to face the daunting averaging problem
well-known from cosmology~\cite{AveragingNonPert,Averaging}.
\end{description}
While reduction is a~standard classical procedure, selection or projection cannot be performed by current
means.
In quantum cosmology\footnote{A great danger in quantum cosmology is that its procedures amount to neither
reduction nor selection nor projection, but rather to {\em production}~-- not a~combination of {\em
pro}jection and re{\em duction} as the word might suggest, but just the presentation of an artif\/icial
model of unknown pedigree.}, the best one may attempt is therefore a~quantum version of reduction, going
beyond pure minisuperspace quantizations but not directly accessing the full solution space.
This is the topic of the present article.

\subsection{Quantum cosmology}

Several dif\/f\/iculties arise when one tries to extend classical reductions to quantum theory.
Unlike classically, symmetric solutions can no longer be exact: Inhomogeneous degrees of freedom are set to
zero, both for conf\/iguration and momentum variables.
The spatial metric at any time must be invariant under the given symmetry, and so do its rate of change or
extrinsic curvature for the metric to be able to remain invariant.
Setting non-symmetric modes of both canonical f\/ields to zero violates the uncertainty principle, and
symmetric quantum solutions cannot be exact solutions of the full theory.
Reduced quantum models can at best be approximations, but making sense of the approximation scheme in
a~clear way remains one of the outstanding challenges to be faced.

Mimicking classical constructions, the kinematical structure of quantum gravity can be reduced by making
use of a~mapping $\sigma\colon{\cal H}_{\rm hom}\to {\cal D}_{\rm full}$ from the kinematical Hilbert space
${\cal H}_{\rm hom}$ of a~homogeneous minisuperspace model, quantizing the degrees of freedom $a_I$, to the
space of distributional states in the full theory~\cite{SymmRed}.
This mapping is analogous to the classical ${\cal M}\to {\cal S}$, but the distributional nature of the
target space spells additional complications.
Moreover, with uncertainty relations violated, symmetric quantum evolution is not exact in the full theo\-ry.
Starting with a~homogeneous full (but distributional) state $\psi_{\rm hom}\in\sigma({\cal H}_{\rm
hom})\subset {\cal D}_{\rm full}$, the distributional extension of (the dual action of) the full
constraints $\hat{C}_{\rm full}$ or their gauge f\/lows $\exp(i\delta \hat{C}_{\rm full})$ does not leave
the state in the image of homogeneous states.
In loop quantum cosmology, methods exist to def\/ine and analyze maps $\sigma$,\footnote{More precisely, as
detailed below, instead of ${\cal D}_{\rm full}$ a~distributional space based on lattices, and therefore
fully inhomogeneous but not the most general states, are used.
Otherwise, if edges not adapted to the symmetry appear, there are obstructions to embeddings of
states~\cite{AinvinA}.} but the derivation of a~dynamical f\/low from the full theory remains dif\/f\/icult
even though candidates do exist.
Loop quantum cosmo\-lo\-gy therefore realizes an incomplete quantum reduction.
Wheeler--DeWitt quantum cosmology, on the other hand, is not a~reduction but a~pure minisuperspace
quantization since no analog of $\sigma$ exists.
(There are also models of loop quantum cosmology which do not go beyond pure minisuperspace models,
disregarding proper considerations of $\sigma$.)

In order to restrict or truncate full quantum evolution (or gauge f\/lows) to homogeneous states, one must
specify a~projection of $\hat{C}_{\rm full}\psi_{\rm hom}$ or $\exp(i\delta \hat{C}_{\rm full})\psi_{\rm
hom}$ back to the image of ${\cal H}_{\rm hom}$ in ${\cal D}_{\rm full}$, for all states $\psi_{\rm hom}$
and all full constraints~-- some part of the averaging problem plays a~role even for reduction when
quantum ef\/fects are involved, another indication of more-severe problems.
No such projection has been provided so far, and therefore the dynamics of reduced models, let alone
minisuperspace models, remains incomplete.

The problem is challenging not just owing to quantum issues, such as the distributional nature of symmetric
states.
Even classically, the question of how to project a~non-symmetric metric to a~homogeneous one is
complicated, and unresolved in its generality; it constitutes the averaging problem of cosmology.
Since a~complete derivation of reduced quantum models from the full theory of some form would, in its
semiclassical limit, include a~solution to the averaging problem, one cannot expect progress on the
dynamical front of quantum reduction unless the classical averaging problem is better understood.

The averaging problem remaining open, the only way at present to go beyond minisuperspace models is to use
properties of a~homogeneous background for an implementation of inhomoge\-nei\-ty, perhaps by perturbation
theory.
In classical cosmology, one commonly makes use of this perspective when inhomogeneous f\/ields are expanded
by Fourier transformation with respect to the modes on a~homogeneous background.
Classically, the approximation is well-understood.
In quantum cosmology, the procedure suf\/fers from the same problems encountered for homogeneous models,
and adds new ones related to the quantization of inhomogeneous modes.
Also the question whether results may depend sensitively on the background (and often gauge) chosen before
quantization remains thorny, related to the complicated anomaly issue of quantum gravity.
(Some anomaly-free realizations with partial quantum ef\/fects
exist~\cite{ConstraintAlgebra,LTBII,ScalarHol,TwoPlusOneDef,ModCollapse,ThreeDeform, JR}.
The anomaly problem is to be faced in canonical and covariant approaches alike: in canonical approaches it
appears in the constraint algebra; in covariant ones, in the path-integral measure or in the correct choice
of face amplitudes of spin foams~\cite{Anomaly}.)

Facing these dif\/f\/iculties, it is an ef\/fective viewpoint which allows progress, making use of
suf\/f\/iciently general parameterizations of quantum ef\/fects, but disregarding f\/ine details.
The viewpoint is half-way between minisuperspace models and a~complete dynamical embedding in the full
theory: One avoids the averaging problem by using inhomogeneous model states adapted to the symmetry, much
like classical Fourier modes on a~given background.
In practice, it is often lattice states with links along the symmetry generators that allow one to include
a~discrete version of inhomogeneity at the quantum level~\cite{InhomLattice}.
A background structure is then built into the framework, but it becomes possible to deal with
inhomogeneity, going beyond pure minisuperspace models and escaping their limitations and artefacts.
A background or some gauge f\/ixing has entered, possibly giving rise to new problems.
But at this stage, ef\/fects suf\/f\/iciently general and parameterized can give reliable access to the
physics involved.

In loop quantum cosmology, this procedure has been developed to the degree that cosmo\-lo\-gi\-cal phenomenology
can be done.
The kinematical structure~-- the basic algebra of holonomy and f\/lux operators~-- can be derived from
the full one.
Evolution and the dynamics, facing the classical averaging problem and the anomaly problem of quantum
gravity, remain much less understood, but here parameterizations have been developed that capture
interesting ef\/fects.
Especially the interplay of various quantum corrections, signif\/icant in dif\/ferent regimes of curvature,
puts restrictions on possible phenomena.

We will now go back to the basics of these constructions to clarify and generalize several mathematical
objects involved.
We will point out one major problem due to oft-used Abeliani\-za\-tions of cosmological models, overlooked so
far.
Its solution has several ramif\/ications even at the level of formulating the dynamics.

\section{Loop quantum cosmology}
\label{s:Mini}

Loop quantum gravity is of the type of a~quantized co-tangent bundle of the space of connections, with
additional constraints that restrict solutions to covariant dynamics.
One usually follows Dirac's quantization procedure, in which one f\/irst f\/inds a~representation of
connection components and their conjugate momenta as operators on a~suitable state space, and then requires
physical states to be invariant under the f\/low generated by constraint operators.
The same separation of kinematics (representing the connection and its momentum) and dynamics (implementing
the constraints) appears in homogeneous models.

In order to formulate the relevant expressions, the abstract index notation is useful and common.
The local connection 1-form is then denoted by $A_a^i$ in component form, while its momentum is $E^a_i$.
Indices $a,b,c,\ldots$ refer to the tangent space of the base manifold, while indices $i,j,k,\ldots$ refer
to the Lie algebra of the structure group, in this context SU(2).
(Geometrically, the momentum plays the role of a~densitized triad, or an orthonormal frame whose components
are multiplied with the absolute value of its determinant.) The positions of indices indicate the dual
nature of the f\/ields (such as 1-forms dual to vector f\/ields), and in products with pairs of mutually
dual indices contraction, or the summation over all values of the paired indices, is understood.
For instance, the densitized triad is related to the inverse spatial metric $q^{ab}$ by $E^a_i E^{bi}=
q^{ab} \det q$.
As used here, indices $i,j,k,\ldots$ are raised or lowered by contraction with the Killing metric of the
structure group (or an application of its associated musical morphism).
Indices $a,b,c,\ldots$ could be raised or lowered with the spatial metric $q_{ab}$, but since it depends on
the $E^a_i$, the f\/ield to be quantized, such operations are usually written explicitly.

In order to quantize the theory, one represents the basic geometric f\/ields, connections $A_a^i$ and
densitized triads $E^a_i$, by integrated versions: holonomies (or parallel transports) of the connection
and f\/luxes of the densitized triad.
(A f\/lux is a~surface integration of $E^a_i$, which is well-def\/ined if one uses the Hodge-type duality
of $E^a_i$ to an su(2)-valued 2-form $\epsilon_{abc}E^a_i$.) In homogeneous models, one tries to f\/ind
versions of these quantities that respect some transitive symmetry acting on space.
Instead of all curves for parallel transports and all surfaces for f\/luxes, one is then led to
a~restricted set.

As a~f\/irst step toward symmetry reduction, mathematical theorems are available to classify and construct
dif\/ferent types of connections invariant under the action of some symmetry group~\cite{KobNom}.
When specialized to homogeneity~\cite{cosmoI}, or a~transitive group action on space, a~set of models
equivalent to the usual Bianchi classif\/ication results: For every Bianchi type, there is a~set of three
left-invariant 1-forms $\omega_a^I$, $I=1,2,3$, obtained as standard invariant 1-forms on the corresponding
transitive symmetry group, for instance by expanding the Maurer--Cartan form $\omega_{\rm MC}=\omega^IT_I$
in terms of Lie-algebra generators $T_I$.
These 1-forms serve as a~basis of the space of invariant connections: All invariant connections are $A_a^i=
\tilde{c}^i_I \omega^I_a$ with spatial constants (but time-dependent) $\tilde{c}^i_I$.
Invariant densitized triads take the dual form, $E^a_i= \tilde{p}^{I}_i X_I^a |\det \omega_a^I|$ with
invariant vector f\/ields $X_I^a$ dual to $\omega_a^I$ (that is, $X_I^a\omega_a^J=\delta_I^J$).
This choice of densitized-triad components ensures that $(8\pi\gamma G)^{-1}\int_{\cal V}
\dot{A}_a^iE^a_i{\rm d}^3x= (8\pi\gamma G)^{-1}V_0\dot{\tilde{c}}^i_I \tilde{p}^I_i$, integrating the
symplectic term of an action over some region ${\cal V}$ of coordinate volume $V_0=\int_{\cal V}{\rm d}^3x$.
Up to constant factors, $\tilde{c}^i_I$ and $\tilde{p}^I_i$ are therefore canonically conjugate:
\begin{gather}
\label{Poisson}
\big\{\tilde{c}^i_I,\tilde{p}^J_j\big\}=\frac{8\pi\gamma G}{V_0}\delta^i_j\delta_I^J.
\end{gather}

A Bianchi metric of the given type is $q_{ab}=\big|\det\big(\tilde{p}^K_k\big)\big|
\tilde{p}^i_I\tilde{p}^i_J\omega_a^I\omega_b^J$ with inverse matrices $\tilde{p}^i_I$ of $\tilde{p}^I_i$.
The metric is invariant under rotations $\tilde{p}^I_i\mapsto R_i^j \tilde{p}^I_j$ with $R\in{\rm SO}(3)$,
generated as gauge transformations by the Gauss constraint $\epsilon_{ij}{}^k \tilde{c}^j_I\tilde{p}^I_k$.
The dif\/feomorphism constraint is not relevant for homogeneous models, and the Hamiltonian constraint
is~\cite{cosmoIII}
\begin{gather}
\nonumber
H=-\frac{1}{8\pi G\sqrt{\big|\det\big(\tilde{p}_i^I\big)\big|}}\Bigl(\epsilon_{ijk}C^K_{IJ}\tilde{c}^i_K\tilde{p}
^I_j\tilde{p}^J_k-\tilde{c}^j_I\tilde{c}^k_J\tilde{p}^I_j\tilde{p}^J_k+\tilde{c}^k_I\tilde{c}^j_J\tilde{p}
^I_j\tilde{p}^J_k
\\
\phantom{H=}
+2\big(1+\gamma^{-2}\big)\big(\tilde{c}^j_I-\Gamma^j_I\big)\big(\tilde{c}^k_J-\Gamma^k_J\big)\tilde{p}_j^{[I}\tilde{p}
_k^{J]}\Bigr)
\label{HamHom}
\end{gather}
with the structure constants $C^K_{IJ}$ of the Bianchi group and the spin connection $\Gamma^i_I$,
depending on~$C^K_{IJ}$ and $\tilde{p}^I_i$.
In diagonal models, where $\tilde{c}^i_I=\tilde{c}_{(I)}\delta_I^i$ and
$\tilde{p}_i^I=\tilde{p}^{(I)}\delta^I_i$ (no summation over~$I$), \eqref{HamHom}~reads~\cite{HomCosmo}
\begin{gather}
H=\frac{1}{8\pi G}\Big(\left((c_2\Gamma_3+c_3\Gamma_2-\Gamma_2\Gamma_3)\big(1+\gamma^{-2}
\big)-n^1c_1-\gamma^{-2}c_2c_3\right)\sqrt{|p^2p^3/p^1|}
\nonumber
\\
\phantom{H=} {}+\left((c_1\Gamma_3+c_3\Gamma_1-\Gamma_1\Gamma_3)\big(1+\gamma^{-2}\big)-n^2c_2-\gamma^{-2}
c_1c_3\right)\sqrt{|p^1p^3/p^2|}
\nonumber
\\
\phantom{H=}{}+\left((c_1\Gamma_2+c_2\Gamma_1-\Gamma_1\Gamma_2)\big(1+\gamma^{-2}\big)-n^3c_3-\gamma^{-2}
c_1c_2\right)\sqrt{|p^1p^2/p^3|}\Big)
\label{H}
\end{gather}
with spin-connection components
\begin{gather*}
\Gamma_I=\frac{1}{2}\left(\frac{p^K}{p^J}n^J+\frac{p^J}{p^K}n^K-\frac{p^Jp^K}{(p^I)^2}n^I\right)
\qquad
\text{for}
\qquad
\epsilon_{IJK}=1
\end{gather*}
and coef\/f\/icients $n^I$ of the Bianchi classif\/ication.

For Bianchi I with $C^K_{IJ}=0$, the constraint reduces to
\begin{gather*}
H=-\frac{1}{8\pi\gamma^2G}\frac{\tilde{c}^j_I\tilde{c}^k_J\tilde{p}^I_j\tilde{p}^J_k-\tilde{c}^k_I\tilde{c}
^j_J\tilde{p}^I_j\tilde{p}^J_k}{\sqrt{\big|\det\big(\tilde{p}_i^I\big)\big|}}.
\end{gather*}
In a~spatially f\/lat isotropic model with $\tilde{c}_I^i=\tilde{c}\delta_I^i$ and
$\tilde{p}^I_i=\tilde{p}\delta^I_i$ with $\tilde{c}=\gamma \dot{a}$ and $|\tilde{p}|= a^2/4$~--
see~\cite{cosmoI,LivRev} for the origin of the factor $1/4$~-- this expression reduces correctly to the
gravitational contribution $H=-3\big(4\pi\gamma^2 G\big)^{-1} \tilde{c}^2 \sqrt{|\tilde{p}|}$ of the Friedmann
equation.

\subsection{Abelian artefacts}
\label{s:arte}

\looseness=-1 One of the f\/irst steps of loop quantization consists in replacing connection components
$\tilde{c}_I^i$ with holonomies or exponentials $\exp\big(\tilde{c}_I^i\tau_i\big)\in {\rm SU}(2)$.
The vast majority of investigations in homogeneous loop quantum cosmology, however, deals with Abelian
models in which the original~SU(2) is replaced by~U(1), thanks either to additional isotropy
symmetries~\cite{IsoCosmo} or a~diagonalization assumption~\cite{HomCosmo}.
With isotropy or diagonalization, a~classical invariant connection automatically becomes Abelian, and the
use of U(1) is not ad hoc but required.
However, in quantum cosmology, Abelian structures turn out to allow specif\/ic choices of Hilbert space
representations not possible in non-Abelian ones.
Such quantizations, based essentially on spaces of functions on the Bohr compactif\/ication of the real
line, are therefore in danger of introducing additional artefacts~-- structural properties that cannot be
met in non-Abelian models, let alone the full theory.
For instance, an isotropic connection has the form $A_a^i=\tilde{c}\delta_a^i$ with just one phase-space
component $\tilde{c}$~\cite{IsoCosmo}.
Mimicking matrix elements of holonomies along straight lines such as the edges of the integration cube
${\cal V}$ (or some other f\/ixed set of edges), one f\/irst represents $\tilde{c}$ by U(1)-holonomies
$h=\exp\big(iV_0^{1/3}\tilde{c}\big)$, or $h^n=\exp(inc)$ with an integer U(1)-representation label~$n$ and
$c:=V_0^{1/3}\tilde{c}$.
By spanning a~function space with superpositions of~$h^n$ for all integer~$n$, all continuous functions on
the group~U(1) are realized.

From U(1)-holonomies $h^n$ one can reconstruct the connection component $c$ only modulo $2\pi$.
One gains full control over the connnection if one considers holonomies along all pieces of the edges of
the integration cube of lengths $\ell_0\leq V_0^{1/3}$, such that holonomies $h^{\mu}=\exp(i\mu c)$ with
\mbox{$\mu\in{\mathbb R}$} result, where $\mu$ may be considered as a~product~$\lambda n$ of the fractional edge
length $\lambda= \ell_0/V_0^{1/3}$ with the representation label~$n$.
Allowing for superpositions of all $h^{\mu}$ as an orthonormal basis, the Hilbert space of all integrable
functions on the Bohr compactif\/ication $\overline{{\mathbb R}}_{\rm Bohr}$ of the real line is
obtained~\cite{Bohr}, rather than a~function space on some periodif\/ication of~${\mathbb R}$.

In this procedure, which has become standard, one implicitly makes use of an identity realized for
representations of $\overline{{\mathbb R}}_{\rm Bohr}$ but lacking a~non-Abelian analog.
In the Abelian case, we start with the U(1)-holonomy $\exp(i\lambda c)$ and evaluate it in the
$\rho_n$-representation: $\rho_n(\exp(i\lambda c))=\exp(i\lambda nc)$.
It so happens that this is the same function of $c$ as obtained from $\rho_{\lambda n}(\exp(ic))$, now
using a~representation of $\overline{{\mathbb R}}_{\rm Bohr}$.
Holonomies in the $n$-representation have led us to the f\/irst expression, which is then identif\/ied with
the latter.
Since they agree as functions, one may base Hilbert space constructions on functions on $\overline{{\mathbb
R}}_{\rm Bohr}$.
However, this step is not available for non-Abelian models, in which case there is no relationship between
$\rho_j(\exp(\lambda A))$ and $\rho_{\lambda j}(\exp(A))$ for $A$ in the Lie algebra of the group, usually
SU(2).
The second expression is not even def\/ined unless $\lambda$ is an integer (or a~half-integer if $j$ is
integer), but even then, the two matrices are unrelated.

\looseness=-1
 If functions on $\overline{{\mathbb R}}_{\rm Bohr}$ are used, one must proceed with care to avoid artefacts
in Abelian mo\-dels, a~problem which has not been realized in existing constructions.
Mathematically, one would confuse $\rho_n(\exp(i\lambda c))$ with $\rho_{\lambda n}(\exp(ic))$, which are
identical as functions of $c$ but have dif\/ferent meanings and are elements of dif\/ferent function spaces.
Physically, merging $\lambda$ and $n$ to one number $\mu=\lambda n$, as done when the Bohr
compactif\/ication is used, eliminates important information because the edge length $\lambda$ and
representation label (or geometrical excitation level) $n$ are then indistinguishable.
In operators, however, $\lambda$ and $n$ should play rather dif\/ferent roles according to what is known
from the full theory: Full holonomy operators act on (spin-network) states labeled by graphs with
representations assigned to their edges.
In the connection representation, such a~state can be written as a~function on the space of connections,
obtained by multiplying matrix elements of all parallel transports along the edges, taken in the respective
representations.
A single holonomy operator adds its curve as a~new edge to the graph if it had not been present before, and
it changes the representation if its curve overlaps with one of the original edges.
Since $\lambda$ in the reduction corresponds to the edge length, operators with dif\/ferent~$\lambda$
should change the underlying graph in dif\/ferent ways, while operators with dif\/ferent~$n$ but the same~$\lambda$ change the graph in the same way but modify the labels dif\/ferently.
These dif\/ferent types of actions cannot be modeled faithfully if reduced operators depend only on the
product~$\lambda n$.
In this section, we present a~new quantization of homogeneous models in which~$\lambda$ and $n$ are kept
separate~-- lifting their degeneracy and assigning to them distinct roles~-- in a~way that extends to
non-Abelian models.

\newpage

\subsection{Homogeneous holonomies}

A local homogeneous connection $A_{\rm hom}$ is a~1-form on the translational symmetry group $S$ underlying
some Bianchi model, taking values in the Lie algebra ${\cal L}G$ of the structure group $G$, $G={\rm
SU}(2)$ for gravity in Ashtekar--Barbero variables.
If the symmetry group acts freely, without any isotropy subgroups, there is a~one-to-one
correspondence~\cite{KobNom} between homogeneous connections according to $S$ and linear maps
$\tilde{\phi}\colon {\cal L}S\to {\cal L}G$ (or elements of ${\cal L}S^*\times{\cal L}G$), not required to
be Lie algebra homomorphisms.
Given $\tilde{\phi}$, the corresponding homogeneous connection is the pull-back $A_{\rm hom}=
\tilde{\phi}^* \omega_{\rm MC}$ under $\tilde{\phi}$ of the Maurer--Cartan form, which latter can be
written as $\omega_{\rm MC}=\omega^IT_I$ in terms of left-invariant 1-forms $\omega^I=\omega_a^I{\rm d}x^a$
on $S$ and its generators $T_I$.
The homogeneous connection components introduced before are the coef\/f\/icients in $\tilde{\phi}(T_I)=
\tilde{c}_I^i \tau_i$ with generators $\tau_i$ of ${\cal L}G$ (here, $\tau_j=-\frac{1}{2}i\sigma_j$ in
terms of Pauli matrices).

A minisuperspace model quantizes the components $\tilde{c}_I^i$, or rather the linear maps
$\tilde{\phi}$.
Both ingredients are in one-to-one correspondence, but the additional structure shown by the linear maps is
useful to decide how dif\/ferent quantum numbers, such as $\lambda$ and $n$ in Section~\ref{s:arte}
should be related to properties of~$S$ (space) and~$G$ (internal space).
We will therefore derive a~quantization based on the mathematical structure of $\tilde{\phi}$.
To extract independent degrees of freedom, we f\/ix a~set of generators $T_I$ of ${\cal L}S$ and understand
a~homogeneous $G$-connection for a~given symmetry group $S$ as a~set of maps $\tilde{\phi}_I\colon \langle
T_I\rangle \to {\cal L}G$ with the scaling condition $\tilde{\phi}_I(rX)= r\tilde{\phi}_I(X)$ for all
$r\in{\mathbb R}$.

Following the methods of loop quantum gravity, we quantize connection components in terms of holonomies.
According to the structure of homogeneous connections, we introduce the notion of homogeneous holonomies by
exponentiation~-- maps $h_{\phi}\colon {\cal L}S\to G, h_{\phi}(X)=\exp(\phi(X))$ with the scaling
condition $h_{\phi}(rX)=h_{r\phi}(X)$.
The maps $\phi_I=L_I\tilde{\phi}_I$ used here dif\/fer from $\tilde{\phi}_I$ by factors of~$L_I$, side
lengths of the integration region ${\cal V}$ of volume $V_0=L_1L_2L_3$, assumed cubic (spanned by the three
generators $T_I\cong X_I^a\partial/\partial x^a$ of the $S$-action).
If the sides of ${\cal V}$ are aligned with the three symmetry generators, one may think of $h_{\phi}(T_I)$
as the holonomy along the corresponding side.
This relationship will be made more precise below.

Elements $X\in{\cal L}S$ to which $\phi$ is applied carry information about the edge used to compute
holonomies $h_{\phi}(X)$.
Referring to the Killing metric on ${\cal L}S$, we decompose $X=\lambda v\not=0$ into its norm
$\lambda=|X|$ and the unit vector $v=X/|X|$, corresponding to the coordinate length of the edge and its
direction.
With the scaling condition, we then have $h_{\phi}(X)= \exp(\lambda \phi(v))$.
We can compute all information about $\phi$ from derivatives $\phi(T_I)= {\rm d} h_{\phi}(\lambda T_I)/{\rm
d}\lambda|_{\lambda=0}$.
We are indeed representing the space of all homogeneous connections, not some periodic identif\/ication.

The dependence of homogeneous holonomies on $\lambda=|X|$ will play an important role in our constructions.
If we consider an edge $e_I$ of coordinate length $\ell_I$ along the generator $X_I^a$, the holonomy, in
general a~path-ordered exponential $h_e={\cal P}\exp(\int_e{\rm d}s A_a^i\tau_i\dot{e}^a)$ of the
connection integrated over a~spatial curve $e(s)$, is $h_{e_I}= \exp(\ell_I \tilde{c}^i_I
\tau_i)=\exp(\ell_IL_I^{-1} \phi(T_I))=h_{\phi}(\lambda_IT_I)$ with $\lambda_I=\ell_I/L_I$.
If all edges are contained in the integration region, we always have $\lambda_I\leq 1$.
More generally, we can allow all real values, but for SU(2), given periodicity of the exponential function,
may restrict to $0\leq\lambda_I<4\pi$ without loss of generality.

For a~given connection, the three choices for $I$, or three directions of space, give rise to three
independent SU(2)-elements $h_{\phi}(T_I)$.
For f\/ixed $\lambda_I$, one can therefore describe the space of homogeneous connections in terms of ${\rm
SU}(2)^3$,\footnote{Thanks to homogeneity, each holonomy transforms as $h_{e_I}\mapsto g h_{e_I}g^{-1}$
under an internal gauge transformation, with the same $g\in{\rm SU}(2)$ for all three edges and on all
their endpoints.
These transformations are identical to those obtained in the full theory for three closed loops
intersecting in one 6-valent vertex~\cite{cosmoI}.
One may picture homogeneous spin-network states as such vertices, but with homogeneity, the vertex
corresponds to all of space~-- homogeneous states are distributional and not given by single spin
networks; see Section~\ref{s:dist}.} but the connection used can be reconstructed completely from
the holonomies only if dif\/ferent choices for $\lambda_I$, or curves of dif\/ferent lengths, are
considered.
If the curves and their lengths are f\/ixed, as in the original constructions
of \cite{cosmoI,IsoCosmo,HomCosmo}, only a~certain periodic identif\/ication of the space of connections is
realized.

In order to generalize homogeneous holonomies $h_{\phi}(T_I)$, we consider a~set ${\cal F}$ of functions
$g_{I}\colon {\mathbb R}\times{\cal L}G\to G$ that fulf\/ill $g_I(rL_I,r^{-1}\tilde{c}_I)=
g_I(L_I,\tilde{c}_I)$ for all $r\in{\mathbb R}$ and $I=1,2,3$ (the scaling condition).
In this way, we can drop the reference to particular edges as appropriate for a~minisuperspace model, but,
as demonstrated in what follows, will still be able to distinguish a~length parameter from a~spin label.
The choice $r=1/L_I$ shows that any such function can be written as $g_I(L_I,\tilde{c}_I)=
\tilde{g}_I(L_I\tilde{c}_I)$ with a~function $\tilde{g}_I$ of just one variable $A\in{\cal L}G$.
If $L_I$ and $r$ are f\/ixed, $g_I$ is simply the group exponential; setting $r$ free allows for
dif\/ferent scalings or dif\/ferent sizes $L_I$ of the integration region within one model.

\subsection{Representation}
\label{s:Rep}

For homogeneous models of loop quantum cosmology, we turn the function space based on holonomies into
a~Hilbert space with an action of holonomies and f\/luxes as basic operators, such that their commutator
corresponds to the classical Poisson bracket~\eqref{Poisson}.
One immediate problem caused by the non-Abelian nature of general connections in combination with
homogeneity regards the way of exponentiating connection components to holonomies and obtaining a~closed
basic algebra for $\{\exp(\lambda_Ic_I^i\tau_i),p^J_j\}$.
Once the path-ordering of inhomogeneous holonomies is no longer available, derivatives of
$\exp(\lambda_Ic_I^i\tau_i)$ by $c_J^j$ will produce extra factors of $\tau_j$ between products of
$c_I^i\tau_i$ in a~power-series expansion of the matrix exponential:
\begin{gather}
\label{Exp}
\frac{\partial\exp(\lambda_Ic_I^i\tau_i)}{\partial c_J^j}=\delta_I^J\sum_{n=0}^{\infty}\frac{\lambda_I^n}
{n!}\sum_{k=1}^n\big(c_I^i\tau_i\big)^{k-1}\tau_j\big(c_I^i\tau_i\big)^{n-k},
\end{gather}
but they do not automatically factor into products of exponentials with $\tau_j$ to mimic the full
holonomy-f\/lux algebra.
(While the cotangent bundle $T^*G$ def\/ines a~natural phase space with group-valued conf\/iguration
variables, it does not necessarily model the correct relation to inhomogeneous holonomies.)

For a~closed basic algebra to result, the factors in derivations of basic holonomy-like functions of
$c_I^i$ may have to be re-ordered, but within a~pure minisuperspace model, there is no guideline, no trace
of the path-ordering left by which one could construct a~natural ordering.
By looking more closely at the relation between basic operators in models and the full theory (or at least
extended curves in holonomies), we will be led to one distinguished choice.

\subsubsection{Hilbert space}

We f\/irst construct a~suitable $C^*$-algebra ${\cal A}$ of functions on homogeneous connections, making
use of our generalized homogeneous holonomies $g_I$: We consider a~function $\psi$ on the space of
homogeneous connections as a~function on the domain of def\/inition ${\mathbb R}\times{\cal L}G$ of $g_I$
which factorizes through $g_I$, that is a~function $\psi(L_I,\tilde{c}_I)$ which can be written as
$\bar{\psi}(g_I(L_I,\tilde{c}_I))$ with a~function $\bar{\psi}$ on $G$.
The scaling condition for $g_I$ then translates into an analogous condition for $\psi$.

\looseness=1
If we f\/ix $L_I$, considering $g_I(L_I,\tilde{c}_I)$ simply as an element of $G$, and refer to the
Peter--Weyl theorem, the general dependence on $\tilde{c}_I$ can be realized by superpositions of functions
$\langle m|\rho_j(g_I(L_I,\tilde{c}_I))| n\rangle$ with all irreducible representations $\rho_j$ of $G$ and
elements $|m\rangle$ and $|n\rangle$ of an orthonormal basis of the representation space of $\rho_j$.
Setting $L_I$ free, with $g_I(\cdot,\tilde{c}_I)$ as a~1-parameter family of $G$-elements, a~larger class
of functions is possible.
However, the scaling condition can be realized only if our functions are superpositions of
$\rho_{\lambda,j}(g_I)^m_n:= \langle m|\rho_j(g_I(\lambda L_I,\tilde{c}_I))|n\rangle$ for
$\lambda\in{\mathbb Q}$.
(The restriction to rational as opposed to real $\lambda$ will be motivated later on.
The labels $\lambda$, $j$, $m$ and $n$ may depend on $I$, but we will often suppress the dependence for
notational simplicity.) Note that, perhaps in a~slight abuse of notation, $\rho_{\lambda,j}$ is not
a~representation of the group ${\mathbb R}\times G$.
It does, however, provide elements of a~suitable function space, whose elements are labeled by~$\lambda$,~$j$,~$m$, and~$n$ and on which we can represent holonomies as constructed in what follows.
(The ${\mathbb R}$-factor is related to curve lengths and not part of the structure group, and therefore
should not be expected to be represented in the standard way of gauge theories.)

We multiply two functions $\rho_{\lambda_1,j_1}(g_I)^{m_1}_{n_1}$ and
$\rho_{\lambda_2,j_2}(g_J)^{m_2}_{n_2}$ as follows: If $I\not=J$, we simply take the product function
depending on $g_I$ and $g_J$ as independent variables, thereby generating a~tensor-product space.
If $I=J$, we write $\lambda_1=N_1z$ and $\lambda_2=N_2z$, with integers~$N_1$ and~$N_2$ and~$z$ the largest
rational number that obeys the two relationships (so that~$N_1$ and~$N_2$ are relatively prime), and
def\/ine
\begin{gather}
\rho_{\lambda_1,j_1}(g_I)^{m_1}_{n_1}\!\cdot\!\rho_{\lambda_2,j_2}(g_I)^{m_2}_{n_2}
:=\sum_{h_1,\ldots,h_{N_1-1},k_1,\ldots k_{N_2-1}}\rho_{z,j_1}(g_I)^{m_1}_{h_1}\rho_{z,j_1}(g_I)^{h_1}
_{h_2}\cdots\rho_{z,j_1}(g_I)^{h_{N_1-1}}_{n_1}
\nonumber
\\
\phantom{\rho_{\lambda_1,j_1}(g_I)^{m_1}_{n_1}\!\cdot\!\rho_{\lambda_2,j_2}(g_I)^{m_2}_{n_2}:=}
\times\rho_{z,j_2}(g_I)^{m_2}_{k_1}\rho_{z,j_2}(g_I)^{k_1}_{k_2}\cdots\rho_{z,j_2}(g_I)^{k_{N_2-1}}_{n_2}.
\label{Mult}
\end{gather}
One may decompose the products of matrix elements on the right-hand side into superpositions of irreducible
contributions to the tensor product $\rho_{j_1}^{\otimes N_1}\otimes \rho_{j_2}^{\otimes N_2}$, akin to
a~spin-network decomposition in the full theory~\cite{RS:spinnet}.
The product is then again a~superposition of $\rho_{\lambda,j}(g_I)^m_n$.

Multiplication as def\/ined is commutative and associative because these properties are respected by the
conditions def\/ining $z$.
(For associativity, the condition that $z$ be maximal is crucial.
There is then a~unique $z$ so that $\lambda_i=N_iz$ for any given number $n$ of $\lambda_i$,
$i=1,\ldots,n$.) There is a~unit element given by $\lambda=0$ (in which case the value of $j$ does not
matter).
For va\-nishing $j$, having the trivial representation of $G$, one may expect a~trivial action, too.
However, according to~\eqref{Mult}, multiplication with $\rho_{\lambda,0}(g_I)$ for $\lambda\not=0$ may
still give rise to decompositions of factors, providing a~dif\/ferent form of the function product even
though the values taken by the original function and its product with $\rho_{\lambda,0}(g_I)$ do not
dif\/fer.
The functions $\rho_{\lambda,0}(g_I)$ (mat\-rix indices are not required in the trivial representation) play
the role of ref\/inement operators, decomposing a~holonomy into pieces whose length is determined by
$\lambda$ in relation to the corresponding parameter of the state acted on.
Since $\rho_0(h_{\phi}(\lambda T_I))=1$ classically, these ref\/inement operators have no classical analog,
as one may expect for a~classical theory knowing nothing about the underlying discreteness.

We def\/ine a~star operation by $(\rho_{\lambda,j}(g_I)^m_n)^*:= \overline{\rho_{\lambda,j}(g_I)^m_n}$
(related to matrix elements of the dual representation $\rho_j^*$ of $\rho_j$ for unitary groups).
The space of functions turns into an Abelian $C^*$-algebra with the supremum norm, assuming $G$ to be
compact.
The supremum is obtained by evaluating $\rho_{\lambda,j}(g_I)^m_n$ on ${\cal L}G$.
(Thanks to the scaling condition, the supremum does not depend on $\lambda$.) The $C^*$-identity
$||\rho^*\cdot \rho||= ||\rho^*||\: ||\rho||$ then follows as it does for the standard example of functions
on $G$.

A Hilbert space structure is obtained by combining the product rule with the Haar measure on $G$.
We def\/ine the inner product
\begin{gather}
\label{InnProd}
\big(\rho_{\lambda_1,j_1}(g_I)^{m_1}_{n_1},\rho_{\lambda_2,j_2}(g_J)^{m_2}_{n_2}\big):=\prod_K\int_G{\rm d}
\mu_{\rm H}(g_K(z L_K,\tilde{c}_K))(\rho_{\lambda_1,j_1}(g_I)^{m_1}_{n_1})^*\cdot\rho_{\lambda_2,j_2}
(g_J)^{m_2}_{n_2},\!\!\!
\end{gather}
where $z$ is def\/ined as before for given $\lambda_1$ and $\lambda_2$.
Equation~\eqref{InnProd} is just the standard inner product on $G$ if we realize, using~\eqref{Mult}, that
the relevant degree of freedom of $(\rho_{\lambda_1,j_1}(g_I)^{m_1}_{n_1})^* \cdot
\rho_{\lambda_2,j_2}(g_J)^{m_2}_{n_2}$ is $g_K(zL_K,\tilde{c}_K)$.
On the right-hand side of this equation, the value of $z$ no longer matters because we integrate over all
group elements $g_K(zL_K,\tilde{c}_K)$, the sole arguments of $(\rho_{\lambda_1,j_1}(g_I)^{m_1}_{n_1})^*
\cdot \rho_{\lambda_2,j_2}(g_J)^{m_2}_{n_2}$.

\subsubsection{Flux operators}
\label{s:Flux}

Components $p^I_i=L_JL_K\tilde{p}^I_i$ ($\epsilon_{IJK}=1$) of the densitized triad, canonically conjugate
to $c_I^i=L_I\tilde{c}_I^i$ via $\{c_I^i,p^J_j\}= 8\pi\gamma G \delta_I^J\delta^i_j$, are quantized to
operators with action
\begin{gather}
\label{Deriv}
\hat{p}^I_i\rho_{\lambda,j}(g_J)^m_n:=-8\pi i\gamma\ell_{\rm P}^2\lambda\delta^I_J\rho_{\lambda,j}
(\tau_i g_J)^m_n
\end{gather}
on our Hilbert space, where we def\/ine the short form $\rho_{\lambda,j}(\tau_i g_J)^m_n:= \sum_k
\rho_j(\tau_i)^m_k \rho_{\lambda,j}(g_J)^k_n$.

Non-Abelian f\/lux operators as def\/ined are {\em not} symmetric because the product~\eqref{Mult}
in~\eqref{InnProd} in general includes a~decomposition.
(One may think of the measure factor as including ref\/inement operators $\rho_{z,0}(g_K)$.) While a~f\/lux
operator acting on either entry in the inner product inserts a~$\tau_i$ to the left of $g_J$ according
to~\eqref{Deriv}, the integration required to evaluate the inner product splits holonomies according to the rational number $z$ depending on $\lambda_1$ and $\lambda_2$.
Integration by parts, performed after multiplying according to~\eqref{Mult} and decomposing, would then
insert $\tau_i$ in each decomposed contribution, not just to the left of the whole $g_K$.

A convenient set of states in the homogeneous Hilbert space is given by a~form of spin-network functions,
depending on the connection via f\/initely many holonomies,
\begin{gather*}
\psi(g_1,g_2,g_3)=\sum_{\lambda_J^{(k)},j_J^{(k)},m_J^{(k)},n_J^{(k)}}\psi_{\lambda_1^{(k)},\lambda_2^{(k)},
\lambda_3^{(k)},j_1^{(k)},j_2^{(k)},j_3^{(k)},m_1^{(k)},m_2^{(k)},m_3^{(k)},n_1^{(k)},n_2^{(k)},n_3^{(k)}}
\\
\phantom{\psi(g_1,g_2,g_3)=}
\times
\prod_{I=1}^3\rho_{\lambda_I^{(k)},j_I^{(k)}}(g_I)^{m_I^{(k)}}_{n_I^{(k)}}
\end{gather*}
with coef\/f\/icients
$\psi_{\lambda_1^{(k)},\lambda_2^{(k)},\lambda_3^{(k)},j_1^{(k)},j_2^{(k)},j_3^{(k)},m_1^{(k)},
m_2^{(k)},m_3^{(k)},n_1^{(k)},n_2^{(k)},n_3^{(k)}}$ for $k$ in some f\/inite index set.
Acting on the contribution $\rho_{\lambda_I,j_I}(g_I)$, $\hat{J}^I_i=(8\pi\gamma\ell_{\rm
P}^2)^{-1}\lambda_I^{-1}\hat{p}^I_i$ satisf\/ies the ${\cal L}G$-algebra by the def\/inition
in~\eqref{Deriv}.
For SU(2), as used in loop quantum gravity, the f\/lux spectrum is therefore given by all numbers
$8\pi\gamma\ell_{\rm P}^2\lambda m$ with half-integer $m$, and the area spectrum (or the spectrum of
$\sqrt{\hat{p}^I_i\hat{p}^{(I)i}}$) by $8\pi\gamma\ell_{\rm P}^2\lambda \sqrt{j(j+1)}$.
Although these eigenvalues are real, one can see the non-symmetric nature of non-Abelian f\/lux operators:
Eigenstates with dif\/ferent eigenvalues (specif\/ically, states with the same $j$ but dif\/ferent
$\lambda$) are not necessarily orthogonal, again owing to the decomposition in~\eqref{InnProd}.
With all rational $\lambda$ allowed, these spectra form continuous sets, but all eigenstates are
normalizable.
The spectra are pure point.

\subsubsection{Heuristics and properties}

To summarize so far, the non-Abelian nature of connections requires care in the proper ordering of
holonomy-type variables to be used in lieu of connections for a~loop quantization.
We refer to edge-integrated holonomies rather than pointwise exponentials, even in homogeneous models.
But since the connection is still homogeneous, we must specify the product rule~\eqref{Mult}~-- the
f\/irst place in our constructions where dif\/ferent holonomies may be compared~-- so that the correct
reduction of degrees of freedom is realized.
This requirement leads us to the decomposition rule, which then further motivates def\/initions of
compatible inner products and derivative operators, including non-selfadjoint features of the latter.
(Decomposition is not an issue in Abelian models because they obey $\rho_{\lambda,n}(\exp(iLc))=
\rho_n(\exp(iLc/r))^p$ for $\lambda=p/r$ with integer $p$ and $r$, an identity that trivially brings all
holonomies to the same distance $L$.
However, this feature makes use of the Abelian coincidence discussed in Section~\ref{s:arte}.
See below for more on Abelian models.)

\begin{figure}[t] \centering
\includegraphics[width=5cm]{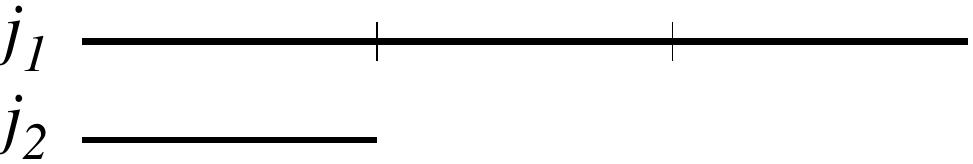} \hspace{3cm} \includegraphics[width=2cm]{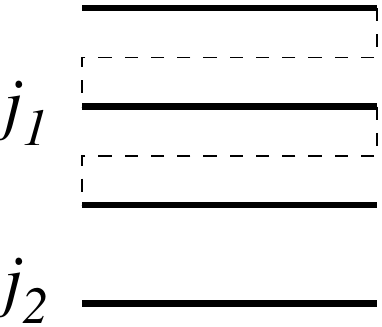}
\caption{Evaluated in homogeneous connections, dif\/ferent pieces of equal length in one holonomy amount to
the same function (left).
To avoid overcounting degrees of freedom, multiplying a~holonomy with a~shorter one requires
a~decomposition into pieces of maximal common length (right), giving rise to the rule~\eqref{Mult}.}
\label{Hol}
\end{figure}

The multiplication rule~\eqref{Mult} observes homogeneity: One can interpret the law as a~decomposition of
two initial holonomies of dif\/ferent lengths as products of equal-length pieces.
Without homogeneity, these pieces at dif\/ferent places would be independent, but homogeneity makes them
identical.
We therefore take the tensor product of all small pieces, split as illustrated in Fig.~\ref{Hol}, and sum
over indices according to the product form.
At this stage, it becomes clear why $\lambda$ should be rational: For incommensurate $\lambda_1$ and
$\lambda_2$, the product $\rho_{\lambda_1,j_1}(g_I)^{m_1}_{n_1}\cdot \rho_{\lambda_2,j_2}(g_I)^{m_2}_{n_2}$
would have to be split into inf\/initely many inf\/initesimally small pieces\footnote{Projective-limit
constructions, in which states with a~given denominator $r$ would play the analog of f\/ixed-state
cylindrical states in the full theory, may be used to allow for incommensurate parameters $\lambda$, but we
will not need this in the present article.}.

In this picture, non-symmetric f\/lux operators~\eqref{Deriv} arise because reduction entails averaging and
decomposition, and what may appear as a~simple quantization of the densitized triad in a~pure
minisuperspace model turns out to be a~more complicated operator when inhomogeneous degrees of freedom are
taken into account.
We will see this in more detail in Section~\ref{s:AvOp}.
Even non-symmetric f\/lux operators (quantizing the densitized triad combined with the action of
holonomy-dependent decomposition, realized for instance by $\rho_{\lambda,0}(g_I)$) model the behavior of
the full theory, in which f\/luxes are self-adjoint.

Also the commutator $[\rho_{\lambda,j}(g_J)^m_n,\hat{p}^I_i]$ of basic operators in the non-Abelian
holonomy-f\/lux algebra requires care due to decomposition.
Up to ordering, it equals $8\pi i\gamma \ell_{\rm P}^2 \lambda \delta_J^I \rho_{\lambda,j}(\tau_i g_J)^m_n$
and corresponds to $i\hbar$ times the Poisson bracket of $\rho_j(\exp(\lambda c_J^i\tau_i))^m_n$, the
classical analog of $\rho_{\lambda,j}(g_J)^m_n$, and $p_i^I$.
For an example of ordering issues, as indicated in the beginning of Section~\ref{s:Rep}, look at the
commutator $[\rho_{1/2,j}(g_I)^M_N,\hat{p}^I_i]$ acting on the state $\psi(g_I)=\rho_{1,j}(g_I)^m_n$.
Acting with $\hat{p}^I_i$ f\/irst produces a~single insertion of $\tau_i$ to the left of $g_I$.
Acting with $\hat{p}^I_i$ after the action of $\rho_{1/2,j}(g_I)$ produces two insertions because we
f\/irst decompose $\rho_{1/2,j}(g_I)^{m'}_{n'}\cdot
\rho_{1,j}(g_I)^m_n=\sum_k\rho_{1/2,j}(g_I)^{m'}_{n'}\rho_{1/2,j}(g_I)^m_k \rho_{1/2,j}(g_I)^k_n$
according to~\eqref{Mult} and use the Leibniz rule.
The result,
\begin{gather*}
4\pi i\gamma\ell_{\rm P}^2\rho_{1/2,j}(\tau_i g_I)^M_N\psi(g_I)+4\pi i\gamma\ell_{\rm P}^2\rho_{1/2,j}(g_I)^M_N
\!
\left(\!\sum_k\rho_{1/2,j}(g_J)^m_k\rho_{1/2,j}(\tau_ig_I)^k_n-\rho_{1,j}(\tau_i g_I)^m_n\!\right)
\end{gather*}
is as expected up to the second term, a~contribution that can be made to vanish by reordering.

Similar calculations for arbitrary $\lambda_1$ and $\lambda_2$ in
$[\rho_{\lambda_1,j_1}(g_I)^M_N,\hat{p}^I_i]$ acting on a~state $\psi(g_I)=\rho_{\lambda_2,j_2}(g_I)^m_n$
($J$-contributions with $J\not=I$ do not matter) result in
\begin{gather*}
[\rho_{\lambda,j}(g_I)^M_N,\hat{p}^I_i]=8\pi i\gamma\ell_{\rm P}^2\lambda\rho_{\lambda,j}
(\tau_ig_I)^M_N+8\pi i\gamma\ell_{\rm P}^2\hat{R}_i^I\rho_{\lambda,j}(g_I)^M_N.
\end{gather*}
The specif\/ic ordering introduces an extra contribution from the reordering operator $\hat{R}_i^I$
def\/i\-ned~by
\begin{gather*}
\hat{R}_i^I\bigl(\rho_{\lambda_1,j_1}(g_I)^{m_1}_{n_1}\cdot
\cdots
\cdot\rho_{\lambda_N,j_N}(g_I)^{m_N}
_{n_N}\bigr)=z\sum_{n=1}^N\rho_{\lambda_1,j_1}(g_I) \cdots
\\
\qquad{}
\times\left(\sum_{p=1}^{N_n}\sum_{k_1,\ldots,k_{N_n-1}}\rho_{z,j_n}(g_I)^{m_n}_{k_1}\cdots\rho_{z,j_n}
(\tau_i g_I)^{k_p}_{k_{p+1}}\cdots\rho_{z,j_n}(g_I)^{k_{N_n-1}}_{n_n}\right)
 \cdots \rho_{\lambda_N,j_N}(g_I)
\\
\qquad
{}-\sum_{n=1}^N\lambda_n\rho_{\lambda_1,j_1}(g_I)^{m_1}_{n_1} \cdots \rho_{\lambda_n,j_n}
(\tau_ig_I)^{m_n}_{n_n} \cdots \rho_{\lambda_N,j_N}(g_I)^{m_N}_{n_N}
\end{gather*}
with $z$ def\/ined as before but for all $\lambda_n$ involved: $\lambda_n=N_nz$ with integers $N_n$ and
ra\-tio\-nal~$z$ maximal.
Up to ordering, $\hat{R}_i^I(\rho_{\lambda_1,j_1}(g_I)^{m_1}_{n_1}\cdots
\rho_{\lambda_N,j_N}(g_I)^{m_N}_{n_N})$ vanishes.
For more compact notation, one may express the ref\/inement included in the action of $\hat{R}_i^I$ in
terms of $\rho_{\lambda,0}(g_I)$ with suitable $\lambda$.
Therefore, $\hat{R}_i^I$ is not independent of the operators already introduced.
Specif\/ically, we can write $\hat{R}_i^I=-(8\pi i \gamma\ell_{\rm P}^2)^{-1} \hat{p}^I_i
(\rho_{z,0}(g_I)-1)$ where $z$ is determined as before for all $\lambda$-parameters of holonomy factors to
the right of $\rho_{z,0}(g_I)$.
The basic commutator then reads
\begin{gather*}
\big[\rho_{\lambda,j}(g_I)^M_N,\hat{p}^I_i\big]=-\big(\hat{p}^I_i\rho_{\lambda,j}(g_I)^M_N\big)-\hat{p}^I_i(\rho_{z,0}
(g_I)-1)\rho_{\lambda,j}(g_I)^M_N=\widehat{\big\{\rho_j(h_{\phi}(\lambda T_I))^M_N,p^I_i\big\}}.
\end{gather*}
Since $\rho_{\lambda,0}(g_I)=1$ classically, the commutator can indeed play the role of a~quantization of
the Poisson bracket, $\widehat{\big\{\rho_j(h_{\phi}(\lambda T_I))^M_N,p^I_i\big\}}= 8\pi i\gamma\ell_{\rm P}^2
\lambda \rho_{\lambda,j}(\tau_ig_I)^M_N$.
Note that it does not quantize~\eqref{Exp}.
In addition to a~quantum representation, we have to choose an ordering of non-Abelian terms when we
def\/ine~\eqref{Deriv}, corresponding to our realization of connection degrees of freedom.
We do not represent pointwise exponentials, which as argued after~\eqref{Exp} would not give rise to
a~closed algebra, but rather, as we will conf\/irm later, use re-ordered classical exponentials closer to
integrated holonomies.
The basic algebra is more complicated than one may have expected, but we do obtain a~closed algebra of
basic operators with the correct classical limit.

Each operator $\hat{p}_j^I$ can be viewed as a~component of the f\/lux operator for a~surface given by the
full side of a~cubic ${\cal V}$, transversal to the direction $X_I^a$.
These minisuperspace f\/luxes refer just to the artif\/icial integration region ${\cal V}$ and its
coordinate volume $V_0=L_1L_2L_3$, not to actual edge lengths or areas of dual surfaces, for instance in
a~lattice~-- we are still quantizing homogeneous connections.
One can easily rescale the linear f\/lux by using $\lambda_J\lambda_Kp_i^I= \ell_J\ell_K \tilde{p}_i^I$ for
$\epsilon_{IJK}=1$, now corresponding to the f\/lux through a~surface of area related to edge lengths
$\ell_J=\lambda_JL_J$ and $\ell_K=\lambda_KL_K$.
Without a~lattice construction to show how dual surfaces are related to links, however, a~strict
minisuperspace quantization does not provide a~satisfactory, ${\cal V}$-independent def\/inition of
f\/luxes.
We will complete the construction in Section~\ref{s:Aver} when considering the relation to the full
theory.

The construction provided here is certainly more complicated than a~traditional minisuperspace
quantization, but it has several convenient features:
\begin{itemize}\itemsep=0pt
\item Because $\phi$ is required to be a~linear map, we need consider holonomies only along linear
combinations of the generators $T_I$, identif\/ied with tangent vectors along ``straight lines''. No
path-ordering is required to compute these holonomies, and they have simple forms.
(For examples of more complicated holonomies, see~\cite{AinvinA}.) In particular, their matrix elements
span a~non-Abelian version of the space of almost-periodic functions: superpositions of periodic functions
with dif\/ferent periodicities given by $\lambda$.
\item The space of homogeneous holonomies is compact if $G$ is compact, as the spectrum of a~unital Abelian
$C^*$-algebra.
Homogeneous loop quantum cosmology makes use of function spaces on compactif\/ications of the classical
spaces of homogeneous connections, which contain all classical connections as a~dense subset.
\item If $G$ is Abelian, the homogeneous Hilbert space is (non-unitarily) related to a~product of the
Hilbert space ${\cal H}_{\rm Bohr}$ of square-integrable functions on the Bohr compactif\/ication of the
real line, with $\dim(S)\times\dim(G)$ factors.
More details will be given below.
\item If $S$ does not act freely, its isotropy subgroup requires additional identif\/ications of the
components of $\tilde{\phi}_I^i$: Linear maps $\tilde{\phi}$ must then satisfy $\tilde{\phi}\circ {\rm
ad}_f= {\rm ad}_{F(f)}\tilde{\phi}$ for any $f$ in the isotropy subgroup and a~corresponding element
$F(f)\in G$; see~\cite{CUP,SymmRed, Brodbeck} for details.
Accordingly, we restrict homogeneous holonomies by $h_{\phi}({\rm ad}_fX)= h_{{\rm ad}_{F(f)} \circ
\phi}(X)$ in addition to the scaling condition.
These functions, and therefore~$g_I$, take values in a~subgroup of~$G$, the centralizer of the subset of
all~$F(f)$ in~$G$, which is Abelian if the isotropy subgroup is suf\/f\/iciently large\footnote{Formally,
classical Abelianization may also be implemented via second-class
constraints~\cite{LQCGaugeFix2, LQCGaugeFix}.}.
\end{itemize}

\subsubsection{Diagonalization}

Detailed ordering prescriptions make some formulas of representations of non-Abelian models look rather
complicated.
It is more straightforward to describe the space of connections by holonomies when diagonal Bianchi models
are used, implying Abelianization.
In diagonal mo\-dels, the relevant parameters appear by writing a~homogeneous connection as
$A_a^i=\tilde{c}_{(I)} \omega_a^I \Lambda_I^i$ with left-invariant 1-forms $\omega_a^I$ of the given
symmetry type and SO(3)-matrices $\Lambda_I^i$ which can easily be f\/ixed to equal $\delta_I^i$.
(Writing $\tilde{c}_I^i= \tilde{c}_{(I)}\Lambda_I^i$, not summing over $I$, makes use of the polar
decomposition of matrices~\cite{HomCosmo}.) The conjugate f\/ield, the densitized triad, has a~similar
decomposition, $E^a_i= \tilde{p}^{(I)} X_I^a \Lambda^I_i |\det \omega_a^I|$.
All ingredients except $\tilde{c}_I$ and $\tilde{p}^I$ are determined by the symmetry type or gauge
choices, and $\tilde{c}_I$ and $\tilde{p}^I$ are the canonical degrees of freedom.
As before, the symplectic term $(8\pi \gamma G)^{-1}\int_{{\cal V}}{\rm d}^3x
\dot{A}_a^iE^a_i=V_0(8\pi\gamma G)^{-1} \dot{\tilde{c}}_I\tilde{p}^I$, integrated over some bounded region
${\cal V}$ of volume $V_0$, provides Poisson brackets
\begin{gather}
\label{PoissonAbel}
\big\{\tilde{c}_I,\tilde{p}^J\big\}=\frac{8\pi\gamma G}{V_0}\delta^J_I.
\end{gather}

Holonomies $h_{e_I}=\exp(\ell_I\tilde{c}_{(I)}\Lambda_I^i\tau_i)$ of diagonal connections still take values
in SU(2).
However, the relations between $h_{e_I}$ for dif\/ferent $I$ are not arbitrary because their generators
$Y_I:=\ell_I\tilde{c}_{(I)}\Lambda_I^i\tau_i$ satisfy the condition ${\rm tr}(Y_IY_J)=0$ for $I\not=J$.
(Dif\/ferent rows or columns of the SO(3)-matrix $\Lambda_I^i$ are orthogonal with respect to the
SU(2)-Killing metric $\eta_{ij}= -2{\rm tr}(\tau_i\tau_j)=\delta_{ij}$.) Although the $h_{e_I}$ do not
commute with one another, any pair of them obeys $gh=hg+h^{-1}g+hg^{-1}- {\rm tr}(hg)$~\cite{HomCosmo}: A
product of diagonal holonomies with $h$ appearing to the right of $g$ can be reordered so that $h$ appears
on the left in all terms.
The gauge structure of diagonal models is essentially Abelian.

To make Abelianization manifest, one usually works with matrix elements $h_I= \exp(i\ell_I
\tilde{c}_I/2)\!\in\! {\rm U}(1)$, completing the reduction of the theory to an Abelian one.
For diagonal models, all phase-space information is indeed captured by these matrix elements because
$\exp(\ell_I\tilde{c}_{(I)} \Lambda_I^i\tau_i)= \cos(\ell_I\tilde{c}_I/2)+2\Lambda^i_I\tau_i
\sin(\ell_I\tilde{c}_{(I)}/2)$.
Fluxes computed for surfaces normal to $X_K^a$ are $F^K=\ell_I\ell_J\tilde{p}^K$ ($\epsilon_{IJK}=1$).
Holonomies as multiplication operators on states in the connection representation and f\/luxes as
derivatives simplify signif\/icantly by Abelianization.
For instance, the volume ope\-ra\-tor, notorious for its complicated spectrum in the full
theory~\cite{VolSpecI,VolSpecII} and also on the 6-valent vertices of non-diagonal homogeneous
models~\cite{cosmoII}, is a~simple product $\hat{V}=\sqrt{|\hat{F}^1\hat{F}^2\hat{F}^3|}$ of derivative
operators $\hat{F}^K=-8\pi i\gamma\lP^2 \lambda_I\lambda_J\partial/\partial c_K$ ($\epsilon_{IJK}=1$) on
U(1).
Abelianization also implies that a~triad representation becomes available.

\subsubsection{Abelian homogeneous connections}

{\sloppy In diagonalized or isotropic models we encounter Abelian homogeneous connections with \mbox{$G={\rm U}(1)$}.
In this case, the structures introduced for general homogeneous connections simplify: Our function space
consists of superpositions of functions $\rho_{\lambda,n}(g)$ of a~single variable
$g(L,\tilde{c})=\exp(iL\tilde{c})$ per independent direction, with $\lambda\in{\mathbb Q}$ and
$n\in{\mathbb N}$, using U(1)-representations \mbox{$\rho_n(g)=g^n$}.
Multiplication~\eqref{Mult} now reads
\begin{gather}
\label{MultAbel}
\rho_{\lambda_1,n_1}(g)\cdot\rho_{\lambda_2,n_2}(g)=\rho_{z,n_1}(g)^{N_1}\rho_{z,n_2}(g)^{N_2}
=\rho_{z,N_1n_1+N_2n_2}(g),
\end{gather}
again with integers $N_1$, $N_2$ and maximal $z$ so that $\lambda_1=N_1z$ and $\lambda_2=N_2z$.
The star relation is $\rho_{\lambda,n}(g)^*= \rho_{\lambda,-n}(g)$, and the inner product~\eqref{InnProd}
evaluates to
\begin{gather}
\nonumber
(\rho_{\lambda_1,n_1}(g),\rho_{\lambda_2,n_2}(g))=\frac{1}{2\pi}\int_0^{2\pi/z}{\rm d}
(zc)\rho_{\lambda_1,-n_1}(\exp(ic))\cdot\rho_{\lambda_2,n_2}(\exp(ic))
\\
\qquad=\frac{1}{2\pi}\int_0^{2\pi}{\rm d}x\rho_{-N_1n_1+N_2n_2}(\exp(ix))=\delta_{N_1n_1,N_2n_2}=
\begin{cases}
0,&\lambda_1n_1\not=\lambda_2n_2,
\\
1,&\lambda_1n_1=\lambda_2n_2
\end{cases}
\label{InnProdAbel}
\end{gather}
after substitution.
Finally, the derivative operator~\eqref{Deriv} is
\begin{gather*}
\hat{p}\rho_{\lambda,n}(g)=8\pi\gamma\ell_{\rm P}^2\lambda n\rho_{\lambda,n}(g)
\end{gather*}
with eigenvalues $8\pi\gamma \ell_{\rm P}^2 \lambda n$.
(This Abelian derivative operator is self-adjoint.)

}

These equations bear some semblance with representations on function spaces on the Bohr compactif\/ication
of the real line, but they are not identical.

\subsubsection{Relation to the Bohr compactif\/ication of the real line}

As recalled in Section~\ref{s:arte}, traditional loop-based minisuperspace quantizations, for
instance in isotropic models, combine the length parameter $\ell_0$ of edges with discrete representation
labels as one real number, giving exponentials $h(\tilde{c})^{n}=\exp(in\ell_0\tilde{c})=\exp(in\lambda c)$
with $c=V_0^{1/3}\tilde{c}$, $\lambda=\ell_0/V_0^{1/3}$, and $\mu=n\lambda\in{\mathbb R}$.
In this Abelian case, homogeneous connections are often viewed as elements of the Bohr compactif\/ication
$\overline{{\mathbb R}}_{\rm Bohr}$ of the real line~\cite{Bohr} (rather than U(1), which is obtained for
f\/ixed $\ell_0$~\cite{IsoCosmo}).
The Bohr compactif\/ication of the real line is a~compact Abelian group with representations in one-to-one
correspondence with those of ${\mathbb R}$: they are given by $z\mapsto z^{\mu}$ for all real
$\mu$.\footnote{Starting from U(1) instead of ${\mathbb R}$, we make the family of representations
continuous by enlarging the group manifold while keeping it compact.
This procedure has no analog for the non-Abelian SU(2), in which non-trivial Lie brackets determine the
representations and discrete spectra of its generators, as well-known from angular momentum in quantum
mechanics.} Functions on $\overline{{\mathbb R}}_{\rm Bohr}$ form a~Hilbert space using the Haar measure
\begin{gather}
\label{BohrInt}
\int{\rm d}\mu_{\rm H}(c)=\lim_{C\to\infty}\frac{1}{2C}\int_{-C}^C{\rm d}c.
\end{gather}
As per the Peter--Weyl theorem, all continuous functions on $\overline{{\mathbb R}}_{\rm Bohr}$ can be
written as countable superpositions
\begin{gather*}
\psi(c)=\sum_{\mu}\psi_{\mu}\exp(i\mu c),
\end{gather*}
states $\exp(i\mu c)$ forming an orthonormal basis for all real $\mu$.
These are eigenstates of the derivative operator $\hat{p}=-8\pi i\gamma\ell_{\rm P}^2{\rm d}/{\rm d}c$ with
eigenvalues $8\pi \gamma\ell_{\rm P}^2\mu$.
(See~\cite{Fewster2008, Velhinho,Velhinho2} for more details on the Bohr compactif\/ication of the real line.)

The form of states suggests a~map between the spaces of functions on Abelian homogeneous connections and
functions on the Bohr compactif\/ication of the real line: $B\colon \rho_{\lambda,n}(g)\mapsto
\exp(i\lambda n c)$ onto the subspace spanned by all $\exp(i\mu c)$ with rational $\mu$.
With the formulas for inner pro\-ducts,~\eqref{InnProdAbel} and~\eqref{BohrInt}, it follows that this map is
an isometry, and it is a~$*$-algebra morphism and commutes with the action of $\hat{p}$.
However, it is not invertible, and therefore not unitary: one can easily f\/ind
$(\lambda_1,n_1)\not=(\lambda_2,n_2)$ such that $\lambda_1n_1=\lambda_2n_2$.\footnote{There is a~bijection
between suitable subspaces of the Bohr--Hilbert space and the Abelian homogeneous Hilbert space with
structure group $G={\rm U}(1)$.
If we take the subspace restricted by $0\leq\lambda<1$, the map $\rho_{\lambda,n}(g)\mapsto
|\mu\rangle:=|\lambda+n\rangle$ is a~one-to-one transformation to the subspace of the Bohr--Hilbert space
with rational $\mu$.
With the restriction on $\lambda$, one can, given $\mu$, uniquely determine $n$ as the integer part of
$\mu$ and $\lambda$ as $\mu-n$.
Moreover, if $\lambda_1+n_1\not=\lambda_2+n_2$, $\lambda_1\not=\lambda_2$ or $n_1\not=n_2$.
However, choices $(\lambda_1,n_1,\lambda_2,n_2)$ exist for which $\lambda_1+n_1\not=\lambda_2+n_2$ but
$\lambda_1n_1=\lambda_2n_2$; the inner product is therefore not preserved and the map is not unitary.}

Not all features of the Bohr compactif\/ication are realized in homogeneous models even of Abelian type;
care is therefore required if only the Bohr compactif\/ication is used:
\begin{enumerate}\itemsep=0pt
\item The label $\mu$ is a~degenerate version of a~pair $(\lambda,n)$ of state parameters, playing distinct
roles in holonomies and discrete dynamics.
The degeneracy of $\lambda$ and $n$ in $\mu$ is lifted by a~direct quantization of homogeneous connections
as linear maps $\phi\colon {\cal L}S\to {\cal L}G$.
\item Our new quantization of homogeneous connections easily applies to non-Abelian models, while the Bohr
compactif\/ication of ${\mathbb R}^3$ does not properly display non-Abelian features of general anisotropic
models.
Via the spectrum of our $C^*$-algebra, we obtain a~compacti\-f\/i\-ca\-tion of the space of non-Abelian
connections unrelated to the Bohr compactif\/ication.
\end{enumerate}

The Bohr compactif\/iction was introduced to loop quantum cosmology in~\cite{Bohr} by way of a~pure
minisuperspace quantization of the isotropic connection component $c$.
Compared to using a~periodif\/ication of the real line to U(1), as originally done in~\cite{IsoCosmo}, this
procedure has the advantage of faithfully representing all values of the connection component: $c$ can be
computed if exponentials $\exp(i\mu c)$ are known for all real $\mu$ (irreducible representations of
$\overline{\mathbb R}_{\rm Bohr}$), while knowing $\exp(inc)$ with integer $n$ (irreducible representations
of U(1)), allows one to compute $c$ only up to adding integer multiples of $2\pi$.
Still, this alteration of the original quantization is inadequate, as shown here.
An isotropic connection is not a~number $c$, and a~diagonal homogeneous connection is not a~triple of
numbers $c_I$, just as an inhomogeneous connection is not a~collection of scalar f\/ields.
A homogeneous connection is a~linear map from ${\cal L}S$ to ${\cal L}G$, or an element of ${\cal
L}S^*\times {\cal L}G$.
The factor of ${\cal L}S^*$ is crucial to relate the nature of a~connection as a~1-form, but it is
overlooked if one takes only the components $c_I$, or a~single $c$ for isotropic models.
The new quantization of homogeneous models provided here takes into account the correct mathematical
structure of homogeneous connections, leading to inequivalent Hilbert space representations.
In some of the following sections, we will see that these dif\/ferences are crucial for realizing
a~relation to the full theory and for some dynamical aspects.

\subsection{Minisuperspace operators and averaging}
\label{s:Aver}

Minisuperspace quantizations allow a~large set of choices regarding quantum representations, kinematical
operators, and, most of all, the dynamics.
The dynamics is the most dif\/f\/icult to derive from the full theory, requiring detailed projection maps
to ensure that one stays on the space of homogeneous states; no strict derivation is available as of now.
Fortunately, however, quantum geometry implies several general ef\/fects in the dynamics, for instance in
Hamiltonian constraint operators of loop quantum cosmology, deviating from classical expressions by the use
of holonomies and inverse-triad operators.
The form of holonomies and inverse triads, in turn, is dictated by properties of the kinematical quantum
representation used.
If one can derive the simpler setting of kinematical representations and basic operators, properties that
imply characteristic dynamics in the full theory are realized in reduced models as well.
Reliable qualitative ef\/fects can be predicted even if the dynamics is not directly derived but rather
constructed by analogy with the full theory, using reduced operators.
Given that the full dynamics so far appears to be ambiguous, too, only generic ef\/fects are reliable,
anyway.
Details of the reduction of dynamics may not matter much, provided one is asking the right questions.
Relating models to the full theory helps one decide which questions can (and should) be asked.

\subsubsection{Lattice subalgebras and spin-lattice states}

For any f\/ixed triple of integers ${\cal N}_I$, the operators $\rho_{k_I/{\cal
N}_I,j_I}(g_I)^{m_I}_{n_I}$, for all integer $k_I$, together with $\hat{p}^J_i$ form a~subalgebra of the
homogeneous holonomy-f\/lux algebra, which we call a~lattice subalgebra or, more specif\/ically, the
$(1/{\cal N}_1,1/{\cal N}_2,1/{\cal N}_3)$-lattice subalgebra.
Any state $\rho_{k_I/{\cal N}_I,j_I}(g_I)^{m_I}_{n_I}|0\rangle$, obtained by acting with a~homogeneous
lattice-subalgebra holonomy on the cyclic state $|0\rangle$ independent of $g_I$, can be written as
a~superposition
\begin{gather*}
\rho_{k_I/{\cal N}_I,j_I}(g_I)^{m_I}_{n_I}=\sum_{h_1,\ldots,h_{k_I-1}}\rho_{1/{\cal N}_I,j_I}(g_I)^{m_I}
_{h_1}\rho_{1/{\cal N}_I,j_I}(g_I)^{h_1}_{h_2}\cdots\rho_{1/{\cal N}_I,j_I}(g_I)^{h_{k_I-1}}_{n_I}
\end{gather*}
of products of elementary excitations $\rho_{1/{\cal N}_I,j_I}(g_I)^{m_I}_{n_I}$.
It can be viewed as the evaluation of a~lattice-based spin-network state in a~homogeneous connection~--
a~cylindrical state whose graph is a~lattice with straight edges and regular spacings $\ell_I=L_I/{\cal
N}_I$.

In order to make contact with inhomogeneous states, we use a~spatial lattice of the form just introduced,
with uniform spacing $\ell_I$ in direction $X_I^a$ as measured in coordinates, from links along the three
invariant vector f\/ields $X_I^a$ of a~Bianchi I model.
We require the region ${\cal V}$ of coordinate size $V_0=L_1L_2L_3$ to be suf\/f\/iciently large, to allow
many lattice links of the chosen spacings.
We must restrict attention to Bianchi I at this stage to obtain closed lattices.
For non-Abelian symmetry groups, such as those of Bianchi models other than type~I, dif\/ferent generators
do not form closed square loops by their integral curves, and therefore no lattice can be
constructed\footnote{For two generators $X_1$ and $X_2$, a~single closed loop is obtained if one uses
integral curves of the left-invariant vector f\/ield of $X_1$ and the right-invariant vector f\/ield of
$X_2$, as proposed in~\cite{APSCurved}: left-invariant vector f\/ields commute with right-invariant ones.
However, no complete lattice can be formed from these integral curves in three spatial dimensions: To
generate lattice sites, one would have to f\/ix one type of vector f\/ield, left- or right-invariant, for
each spatial direction.
If $X_1^a$ is taken as left-invariant, $X_2^a$ must be right-invariant for a~closed 2-dimensional lattice
in the $1-2$-surface.
For a~closed lattice in the $1-3$-surface, also $X_3^a$ would have to be right-invariant, but then, with
both $X_2^a$ and $X_3^a$ right-invariant, there is no closed lattice in the $2-3$-surface~-- unless
$X_2^a$ and $X_3^a$ happen to commute.
Lattice constructions based on the interplay of left- and right-invariant vector f\/ields cannot be
performed for all Bianchi types, making those constructions in the available cases (Bianchi I and II)
non-generic.
Attempts at such constructions in anisotropic models show some of the pitfalls of ad-hoc assumptions, as
illustrated by the series~\cite{ImpBianchiI,ImpBianchiII,ImpBianchiIX} of papers where most initial claims
of~\cite{ImpBianchiI}, for instance regarding averaging or a~possible relation to lattice constructions,
had to be withdrawn or weakened in later installments.
Initially simple-looking constructions became more and more contrived.
Instead, it is more general (while still not free of assumptions) to use lattices according to the
kinematical structure of Bianchi~I, and then implement other Bianchi models by suitable curvature terms in
the dynamics~\cite{Spin}.
In this way, all Bianchi class A models can be quantized with one and the same scheme.
One may worry about an inconsistency in using Bianchi~I lattices for other Bianchi models.
However, at the inhomogeneous lattice level, no strict Bianchi models can be realized.
The symmetry type just provides guidelines along the way to consistent dynamics, which can well be realized
for all class~A Bianchi models.}. As seen in Section~\ref{s:Ham}, such Bianchi models can still be
quantized at least as far as their dynamics is concerned: One would refer to Bianchi~I lattices to def\/ine
basic homogeneous variables, and implement the dif\/ferent dynamics by an additional potential in the
Hamiltonian constraint.
Lacking closed lattices, the spatial manifold structure of other Bianchi models cannot be realized in
a~quantum model.
However, this classical ingredient should not be taken too seriously, anyway, because from inhomogeneous
models the ef\/fects of loop quantum gravity are known to modify the classical space-time structure as
a~consequence of a~quantum-corrected hypersurface-deformation algebra~\cite{Action,BohrWigner}.
Lattices can be seen as a~crutch or a~helpful visualization to construct suitable state spaces on which one
can represent operators with the correct classical limit.
Once such state spaces have been obtained, they can be extended to dif\/ferent kinds of dynamics even if
lattices are no longer available.

{\sloppy Fixing an orientation for each of the three directions, we label lattice links by pairs $(v,I)$ of a~vertex
$v$ as the starting point of a~link $e_{v,I}$ in direction $X_I^a$ (as in~\cite{QuantCorrPert}).
For a~connection $A_a^i$ (not assumed homogeneous at this stage), each link gives rise to a~holonomy
$h_{v,I}= {\cal P}\exp(\int_{e_{v,I}} A_a^i\tau_i\dot{e}^a{\rm d} s)$.
We will work with spin-network states of the underlying lattice (not required to be gauge invariant), or
spin-lattice states.
Each link holonomy appears in some irreducible SU(2)-representation with spin $j_{v,I}$.
In the matrix representation $\rho_{j_{v,I}}(h_{v,I})$, we~pick matrix elements $\langle
m_{v,I}|\rho_{j_{v,I}}(h_{v,I})|n_{v,I}\rangle$, with two eigenstates $|m_{v,I}\rangle$ and
$|n_{v,I}\rangle$ of $\rho_{j_{v,I}}(\tau_3)$ (or~any other component).
The function $\langle m_{v,I}|\rho_{j_{v,I}}(h_{v,I})|n_{v,I}\rangle$ is then an eigenstate with
eigenva\-lues~$m_{v,I}$ and~$n_{v,I}$, respectively, of the 3-components of right-invariant and
left-invariant derivatives by~$h_{v,I}$.
Our spin-lattice states are therefore functions $\psi_{(j_{v,I},m_{v,I},n_{v,I})}(h){=}\prod_{v,I}
\langle m_{v,I}|\rho_{j_{v,I}}(h_{v,I})|n_{v,I}\rangle$ depending on the connection via link holonomies,
with an inner product def\/ined as usual by integrating over all $h_{v,I}$ using the Haar
measure~\cite{FuncInt}.
This def\/ines the Hilbert space ${\cal H}_{\rm lattice}$.
For short, we will write these states as $|(j_{v,I},m_{v,I},n_{v,I})\rangle$.
We have the usual action of holonomies and f\/luxes.

}

\subsubsection{Homogeneous distributions}
\label{s:dist}

A homogeneous analog of spin-lattice states, depending on holonomies $h_{\phi}(\lambda_IT_I)=
\exp(\lambda_I \phi(T_I))$, is $\psi_{(\lambda_I, j_I,m_I,n_I)}(g_J)= \prod_I \langle
m_I|\rho_j(h_{\phi}(\lambda_IT_I))|n_I\rangle= \prod_I \langle
m_I|\rho_{\lambda_I,j_I}(g_I)|n_I\rangle$ with $\lambda_I{\cal N}_I$ integer, written for short as
$|(\lambda_I,j_I,m_I,n_I)\rangle$.
There is an additional label $\lambda_I$, replacing the edge or link dependence of inhomogeneous states and
representing the ${\cal L}S$-part $X=\lambda_IT_I$ of a~homogeneous connection $\phi\in {\cal
L}S^*\otimes{\cal L}{\rm SU}(2)$, subject to the scaling condition.
The set of these states is f\/ixed by holonomies in the $(1/{\cal N}_1,1/{\cal N}_2,1/{\cal N}_3)$-lattice
subalgebra, with elementary holonomies acting by multiplication, changing the SU(2)-representations $j_I$
according to recoupling rules, and f\/lux operators having eigenvalues $8\pi\gamma\ell_{\rm P}^2m_I$ (for
right-invariant vector f\/ields) and $8\pi\gamma\ell_{\rm P}^2n_I$ (for left-invariant ones).
No decomposition as in~\eqref{Mult} is required since we have a~f\/ixed common denominator ${\cal N}_I$ for
all holonomies considered in direction $X_I$.

So far, homogeneous and inhomogeneous lattice states are def\/ined separately from each other.
We relate them by introducing a~map $\sigma\colon {\cal H}_{\rm hom}\to {\cal D}_{\rm lattice},
|(\lambda_I,j_I,m_I,n_I)\rangle\mapsto ((\lambda_I,j_I,m_I,n_I)|$ from the homogeneous Hilbert space to
distributions on the lattice Hilbert space.
This map is the key ingredient of quantum symmetry reduction, as described in
Section~\ref{s:Reduction}.
Following~\cite{SymmRed}, we def\/ine homogeneous distributions by their evaluations
\begin{gather}
((\lambda_I,j_I,m_I,n_I)|(j_{v,I},m_{v,I},n_{v,I})\rangle
\nonumber
\\
\qquad
=\langle(\lambda_I,j_I,m_I,n_I)|(j_{v,I},m_{v,I}
,n_{v,I})\rangle|_{h_{v,I}=\exp(\phi(T_I)/{\cal N}_I)}
\label{Dist}
\end{gather}
on all basis states of ${\cal H}_{\rm lattice}$.
On the right-hand side, the inner product is taken in ${\cal H}_{\rm hom}$, with
$|(j_{v,I},m_{v,I},n_{v,I})\rangle|_{h_{v,I}=\exp(\phi(T_I)/{\cal N}_I)}$ obtained by restricting the
connection dependence of the spin-lattice state to homogeneous $\phi$.
The distributional evaluation vanishes unless the representation $j_I$ appears in the tensor product
$\bigotimes_{v,I}j_{v,I}$, and $m_I=\sum_{v,I}m_{v,I}$,
$n_I=\sum_{v,I}n_{v,I}$.\footnote{In~\eqref{Dist}, we restrict to holonomies
$h_{v,I}=\exp(\lambda_I\phi(T_I))$ with $v$-independent $\lambda_I=1/{\cal N}_I$, or a~regular aligned
lattice of uniform link lengths.
At this stage, we could allow irregular lattices with varying $\ell_I(v)$, as long as all links are still
along symmetry generators $\sum_I\lambda_IT_I$.
Dif\/ferent lattice sectors would then contribute to the reduction, and ref\/inement would be necessary in
the multiplication and action of holonomies.
This option will be discussed in more detail below.} The reduction of states depends on the size of the
region ${\cal V}$ via ${\cal N}_I$, just like the classical reduction of the phase space.

\subsubsection{Averaged operators}
\label{s:AvOp}

An operator $\hat{O}$ can be reduced from the lattice theory to the homogeneous Hilbert space if its dual
action f\/ixes the space of homogeneous distributions: If there is a~$|\psi\rangle\in{\cal H}_{\rm hom}$
such that $((\lambda_I,j_I,m_I,n_I)|\hat{O}^{\dagger}|(j_{v,I},m_{v,I},n_{v,I})\rangle=
(\psi|(j_{v,I},m_{v,I},n_{v,I})\rangle$ for all $|(j_{v,I},m_{v,I},n_{v,I})\rangle$, we def\/ine
$\hat{O}|(\lambda_I,j_I,m_I,n_I)\rangle=|\psi\rangle$.
All link holonomies $\langle m_{v,J}|\rho_{j_{v,J}}(h_{v,J})|n_{v,J}\rangle$ along symmetry generators,
taken as multiplication operators, satisfy this condition.
They act on distributional homogeneous states by $\langle m_{v,J}|\rho_{j_{v,J}}(h_{v,J})|n_{v,J}\rangle
\langle g_I|(\lambda_I,j_I,m_I,n_I)\rangle= \rho_{1/{\cal N}_J,j_{v,J}}(g_J)^{m_{v,J}}_{n_{v,J}}\cdot
\rho_{\lambda_I,j_I}(g_I)^{m_I}_{n_I}$, just as in the reduced space of homogeneous states.

Flux operators require additional constructions.
A single lattice f\/lux $\hat{F}_{v,I}$ associated with a~surface dual to link $e_{v,I}$ does not map
a~homogeneous distribution to another such state: Take a~set of states
$|\psi_{v,I}\rangle:=|0,\ldots,0,(1/2,1/2,1/2),0,\ldots,0\rangle$, each with non-zero labels only on one
lattice link $e_{v,I}$.
We have
\begin{gather*}
((1/{\cal N}_I,j_I,m_I,n_I)|\hat{F}_{v,I}|\psi_{v,I}\big\rangle=4\pi\gamma\ell_{\rm P}^2\delta_{j_I,1/2}
\delta_{m_I,1/2}\delta_{n_I,1/2}
\end{gather*}
and $((\lambda_I,j_I,m_I,n_I)|\hat{F}_{v,I}|\psi_{v',I'}\rangle=0$ if $v\not=v'$ or $I\not=I'$.
Therefore,
\begin{gather*}
((1/{\cal N}_I,1/2,1/2,1/2)|\hat{F}_{v,I}|\psi_{v,I}\big\rangle\not
=((1/{\cal N}_I,1/2,1/2,1/2)|\hat{F}_{v,I}|\psi_{v',I}\big\rangle
\qquad
\text{for}
\qquad
v\not=v'.
\end{gather*}
However, we must have $(\Psi|\psi_{v,I}\rangle= (\Psi|\psi_{v',I}\rangle$ for any homogeneous state
$|\Psi\rangle\in{\cal H}_{\rm hom}$ since $\psi_{v,I}|_{h_{w,I}=\exp(\lambda_I\phi(T_I))}=
\psi_{v',I}|_{h_{w,I}=\exp(\lambda_I\phi(T_I))}$.
The state $((1/{\cal N}_I,1/2,1/2,1/2)|$ cannot be contained in a~decomposition of $((1/{\cal
N}_I,j_I,m_I,n_I)|\hat{F}_{v,I}$ in our basis, and we can repeat the arguments with arbitrary values of the
non-zero label in $|\psi_{v,I}\rangle$ to show that no homogeneous state can be contained in the
decomposition.
Therefore, the distribution $((1/{\cal N}_I,j_I,m_I,n_I)|\hat{F}_{v,I}$ cannot be a~superposition of
homogeneous distributional states: f\/lux operators associated with a~single link do not map the space of
homogeneous states to itself.

Even classically, the f\/low $\{\cdot,F_S\}$ generated by a~f\/lux operator is not everywhere tangent to
the submanifold of homogeneous connections and {\em unrestricted} triads in the inhomogeneous phase space,
but it is tangent to the subspace on which both the connection {\em and} the triad are homogeneous.
In the quantized theory, using distributional states, we have ensured states to be restricted to
homogeneous connections, but no such condition has yet been implemented for the densitized triad or
f\/luxes.

Flux operators must be averaged to generate a~f\/low that keeps the space of homogeneous states invariant.
However, non-Abelian gauge properties prevent us from simply adding\linebreak
$\sum_n \int_{S_n} E^a_in_a{\rm
d}^2y$ for a~family of surfaces $S_n$ translated along the generators of the symmetry group.
Instead, we must relate the f\/ibers of the SU(2)-bundle in which $E^a_i$ takes values, using parallel
transport between the $S_n$.
(This problem seems to be related to issues encountered in constructions of a~non-Abelian triad
representation~\cite{FluxRep}.
Here, homogeneity will help us to propose a~solution.)

To describe the specif\/ic construction, we assume an aligned state, consisting only of ho\-lo\-no\-mies
$h_{v,I}$ in the three independent directions but not necessarily forming a~regular lattice.
For an averaged $p^I_i$, we choose families of surfaces $S_{n,I}$ transversal to the symmetry generators
$X^a_I$, such that they have co-normals $n^I_a=\delta^I_a$, layered at regular intervals across the region
${\cal V}$.
Eventually, we will send the number $N$ of surfaces to inf\/inity.
Before doing so, we def\/ine a~gauge-covariant averaging by $\overline{p}^I_i=N^{-1}\sum\limits_{n=1}^N
\int_{S_{n,I}} {\rm ad}_{h_I(y)}(E^a_i(y)n_a^I){\rm d}^2y$ where $h_I(y)$ is the connection-dependent
parallel transport from some base point, chosen for each integral curve in direction~$I$, to a~point~$y$ on
the surface.

The base points will be chosen in a~state-dependent way because the state determines how the connection is
excited, usually in a~discontinuous way at lattice vertices.
We decompose a~state as a~superposition of contributions $\Psi=\psi(h_{w,J},h_{w',K}) \prod_{v_I}
\rho_{j_{v,I}}(h_{v_I,I})$ where the dependence on holonomies along directions $J$ and $K$ will not matter.
The set of all $v_I$ then gives us all vertices where parallel transport in the $I$-direction changes
discontinuously.
We will average with these vertices chosen as base-points, so that only the continuous parts of parallel
transport are taken into account.
We f\/irst decompose surfaces $S_{n,I}=\bigcup_k S_{n,I,k}$ so that each piece $S_{n,I,l}$
intersects at most one edge.
In the action of the f\/lux operator, instead of summing over $k$ we will then be summing over edges
intersecting the surface: We write{\samepage
\begin{gather*}
\frac{1}{N}\sum_{n=1}^N\int_{S_{n,I}}{\rm d}^2y{\rm ad}_{h_I(y)}\big(\hat{E}^a_i(y)n_a^I\big)\Psi
=\frac{1}{N}\sum_{n=1}^N\sum_{e_{v_I,I}\cap S_{n,I}
\not=\varnothing}{\rm ad}_{h_{v_I,I}(v_n)}\big(\hat{F}_{v_I,I}(S_{n,I,k})\big)\Psi
\end{gather*}
with $h_{v_I,I}(v_n)$ the parallel transport from $v_I$ to the intersection point $v_n$ of $e_{v_I,I}$ with
$S_{n,I}$.}

If a~piece of the surface $S_{n,I}$ intersects an edge $e_{v_I,I}$, we choose the base point to be $v_I$
(for a~right-invariant vector f\/ield quantizing the f\/lux, or the other endpoint of $e_{v_I,I}$ for
a~left-invariant one).
The adjoint action of $h_{v_I,I}(v_n)$ in the averaged f\/lux then implies that a~f\/lux operator does not
insert just $\tau_i$ in the holonomy, at the intersection point $\{v_n\}=S_{n,I}\cap e_{v_I,I}$ with the
surface, but $h_{v_I,I}(v_n)^{-1}\tau_i h_{v_I,I}(v_n)$.
For a~single edge $e$, splitting the holonomy $h_e:=h^{(1)}_e(v_n)h^{(2)}_e(v_n)$ in two pieces
$h^{(1)}_e(v_n)$ and $h^{(2)}_e(v_n)$ at an intersection point $v_n$, we thus have
\begin{gather*}
\hat{\overline{p}}{}^I_i\langle m'|\rho_j(h_e)|n'\rangle
=-8\pi i\gamma\ell_{\rm P}^2\lim_{N\to\infty}\frac{1}{N}\sum_{n=1}^N
\big\langle m'|\rho_j\big(h^{(1)}_e(v_n){\rm ad}_{h_e^{(1)}(v_n)}(\tau_i)h_e^{(2)}(v_n)\big)|n'\big\rangle
\\
\phantom{\hat{\overline{p}}{}^I_i\langle m'|\rho_j(h_e)|n'\rangle}
=-8\pi i\gamma\ell_{\rm P}^2\lim_{N\to\infty}\frac{1}{N}\sum_{n=1}^N\langle m'|\rho_j(\tau_i h_{e}
)|n'\rangle\delta_{S_{n,I}\cap e\not=\varnothing},
\end{gather*}
where only terms such that $S_{n,I}\cap e\not=\varnothing$ contribute.
For large $N$, the number of non-zero contributions divided by $N$ approaches $\lambda_e$, the ratio of the
length spanned by $h_e$ relative to $L_I$.
We obtain eigenvalues $8\pi\gamma\ell_{\rm P}^2 \lambda_e m'$.
Not surprisingly, in this homogeneous context the averaged f\/lux does not refer to any point on the edge
where it acts, but it picks up the relative length of the edge by the number of intersection points.
For multiple edges, the f\/lux acts by the product rule.

If all edges involved form a~regular lattice, with the number ${\cal N}_I=L_I/\ell_I$ of lattice links, it
follows that
\begin{gather}
\label{DerivAv}
\hat{\overline{p}}{}^I_i=\frac{1}{{\cal N}_I}\sum_v\hat{F}_{v,I,i}.
\end{gather}
The factor of $1/{\cal N}_I=\ell_I/L_I$ eliminates over-counting by adding f\/luxes of all lattice sites
along direction $I$.
In the other two directions, on the other hand, we sum rather than average because the minisuperspace
$p_i^I$ is def\/ined for a~surface stretching through the region ${\cal V}$, as in~\eqref{Deriv}.
Indeed, heuristically, the eigenvalues of $\hat{\overline{p}}{}^I_i$, $8\pi\gamma\lP^2 {\cal N}_I^{-1}
\sum_v m_{v,I}$ can be written as $L_JL_K$ multiplying the average value of the densitized triad:
$8\pi\gamma\lP^2{\cal N}_I^{-1} \sum_v m_{v,I}= 8\pi\gamma\lP^2{\cal N}_J{\cal N}_K\overline{m_I}=
{\cal N}_J{\cal N}_K \overline{E^I_3}$, where $\overline{m_I}= ({\cal N}_1{\cal N}_2{\cal
N}_3)^{-1}\sum_v m_{v,I}$ is the lattice average, quantizing the average of the plaquette f\/lux
$\int E^I_3 {\rm d}x^J{\rm d}x^K/8\pi\gamma\lP^2$.

\subsubsection{Kinematical quantization commutes with symmetry reduction}
\label{s:comm}

Holonomy operators in ${\cal H}_{\rm hom}$ are directly obtained from their dual action on ${\cal D}_{\rm lattice}$.
It does not matter whether we act with a~holonomy operator f\/irst and then symmetry-reduce, or f\/irst
reduce and then act with the corresponding homogeneous operator:
\begin{gather*}
\sigma(\rho_{1/{\cal N}_J,j_{v,J}}(g_J)|(\lambda_I,j_I,m_I,n_I)\rangle)=\rho_{j_{v,J}}(h_{v,J}
)\sigma(|(\lambda_I,j_I,m_I,n_I)\rangle).
\end{gather*}

After averaging, the same commutation relationship is realized for lattice f\/lux operators.
We have
\begin{gather*}
((1/{\cal N}_I,j_I,m_I,n_I)|\hat{\overline{p}}{}^J_3|(j_{v,I},m_{v,I},n_{v,I})\big\rangle\\
\qquad{}
=\frac{1}{{\cal N}_J}\sum_v ((1/{\cal N}_I,j_I,m_I,n_I)|\hat{F}_{v,J,3}|(j_{v,I},m_{v,I},n_{v,I})\big\rangle
\\
\qquad {} =\frac{8\pi\gamma\lP^2}{{\cal N}_J}\sum_v m_{v,J}\delta_{j_I\in\bigotimes_v j_{v,I}}
\delta_{m_I,\sum_v m_{v,I}}\delta_{n_I,\sum_vn_{v,I}}
\\
\qquad{} =8\pi\gamma\lP^2\lambda_Jm_J((\lambda_I,j_I,m_I,n_I)|(j_{v,I},m_{v,I},n_{v,I})\rangle
\end{gather*}
using the $\delta$-identif\/ications and $\lambda_J=1/{\cal N}_J$ in the last step.
On reduced states, on the other hand, we have $\hat{p}_3^J|(\lambda_I,j_I,m_I,n_I)\rangle= 8\pi\gamma \lP^2
\lambda_Jm_J$ from the right-invariant vector f\/ield~\eqref{Deriv} in the $\lambda_J$-sector.
Comparing these equations, we see that $\sigma(\hat{p}^J_3 |(\lambda_I,j_I,m_I,n_I)\rangle)=
\hat{\overline{p}}{}^J_3 \sigma(|(\lambda_I,j_I,m_I,n_I)\rangle)$, with analogous calculations for other
components of $\hat{p}_i^J$.
(As remarked in Section~\ref{s:Flux}, the non-Abelian $\hat{p}_i^J$ is not a~symmetric operator
unless we are in the lattice setting of f\/ixed $\lambda_I$ for all states involved.
Similarly, $\hat{\overline{p}}{}_i^J$ is not symmetric in this situation due to averaging, in particular
with a~state-dependent choice of base points for parallel transport.)

For basic operators, it does not matter whether we quantize or reduce f\/irst.
We obtain the same representation properties as in the full theory, and the same qualitative
quantum-geometry ef\/fects.
But a~quantitative correspondence is more complicated for composite operators, especially the Hamiltonian
constraint crucial for dynamics.

\subsubsection{Holonomy-f\/lux algebra in reduced models}

The previous calculations have shown how holonomy-f\/lux representations of homogeneous mo\-dels are derived
from the full algebra.
Using the minisuperspace embedding $\sigma$, we obtain basic operators~-- holonomies and f\/luxes~-- from
their action on inhomogeneous lattice states.
Since the holonomy-f\/lux representation of the full theory is unique~\cite{WeylRep, LOST}, the
minisuperspace representation derived here, by restriction of the full algebra to lattices followed by
taking the dual action on homogeneous distributions, enjoys the same distinction for a~given lattice.

So far, we have written all formulas for the general case of non-Abelian homogeneous models.
Using the same techniques of restriction of states and reduction of operators, it is straightforward to
implement diagonalization or isotropy: We restrict states to $\phi_I^i= c_{(I)} \Lambda_I^i$ for diagonal
models, or $\phi_I^i= c\Lambda_I^i$ for isotropic ones.
Flux operators $\hat{\overline{p}}{}^I= \Lambda_{(I)}^i \hat{\overline{p}}{}^I_i$ then leave diagonal
states invariant (while $\Lambda_J^i\hat{\overline{p}}{}^I_i$ with $J\not=I$ would not), and the averaged
$\hat{\overline{p}}=\frac{1}{3} \sum_I\hat{\overline{p}}{}^I$ leaves isotropic states invariant.
These situations are covered in~\cite{InhomLattice}.

\subsection{Dynamics}

From holonomy and f\/lux operators, we construct more complicated ones such as the volume or the
Hamiltonian constraint.
The volume on spin-lattice states can be def\/ined as in the full theory, using
$|\frac{1}{6}\epsilon^{ijk}\epsilon^{IJK}\hat{F}_{v,I,i} \hat{F}_{v,J,j}\hat{F}_{v,K,k}|^{1/2}$, just
restricted to 6-valent vertices as encountered in a~lattice.
The complete spectrum is unknown in the non-Abelian case.
For simpler algebraic relations, we may replace the cubic SU(2)-invariant with a~product of quadratic
invariants, $\hat{V}_v:=\prod\limits_{I=1}^3(\hat{F}_{v,I,i} \hat{F}_{v,I}{}^i)^{1/4}$ with eigenvalues
$(8\pi\gamma)^{3/2} \ell_{\rm P}^3 \prod\limits_{I=1}^3 (j_{v,I}(j_{v,I}+1))^{1/4}$.
In what follows, details and dif\/ferences of these spectra will not play a~major role, and we will make
use of the simpler version.

As part of the formulation of dynamics, we will be interested in reducing the volume operator.
An important question for non-linear combinations of basic operators is whether the average is taken before
or after reduction.
The minisuperspace volume
\[
\prod\limits_{I=1}^3 \big(\hat{\overline{p}}{}^I_i
\hat{\overline{p}}{}^{I,i}\big)^{1/4}=({\cal N}_1{\cal N}_2{\cal N}_3)^{-1/2} \prod\limits_{I=1}^3
\sqrt[4]{\hat{J}_I^2},
\]
 using $\hat{J}_I=\sum_v \hat{F}_{v,I}$ and~\eqref{DerivAv}, has eigenvalues
\begin{gather*}
\frac{(8\pi\gamma)^{3/2}\ell_{\rm P}^3}{\sqrt{{\cal N}_1{\cal N}_2{\cal N}_3}}\prod_{I=1}
^3\sqrt[4]{j_I(j_I+1)}=(8\pi\gamma)^{3/2}\ell_{\rm P}^3\prod_{I=1}^3\sqrt{\lambda_I}\sqrt[4]{j_I(j_I+1)}.
\end{gather*}
The spectrum can be computed using $\hat{p}^I_i$ in~\eqref{Deriv} or $\hat{\overline{p}}{}^I_i$
in~\eqref{DerivAv}, but it does not equal that of the averaged $({\cal N}_1{\cal N}_2{\cal
N}_3)^{-1}\sum_v V_v$, which would be the reduced volume operator.
The distinction is important for the correct size of quantum-geometry ef\/fects, as we will see below.
Pure minisuperspace models make use of $\hat{p}^I_i$ or $\hat{\overline{p}}{}^I_i$; correctly capturing
quantum ef\/fects requires an averaged volume operator.

\subsubsection{Hamiltonian constraint}
\label{s:Ham}

{\sloppy The classical Hamiltonian constraint contains curvature components, to be represented in homogeneous models
by holonomies $\rho_{\lambda,j}(g_I)^m_n$, non-polynomial functions of the connection which dif\/fer from
connection or curvature components by higher-order terms.
In Abelian models, it is easy to see that $\rho_{\lambda,n}(g_I)$ as an operator is not weakly
continuous in $\lambda$ at $\lambda=0$, and it not possible to def\/ine a~connection operator via ${\rm
d}/{\rm d}\lambda|_{\lambda=0}$: The diago\-nal matrix elements $\langle
(\lambda_I',n_I')|\rho_{\lambda,n}(g_I)|(\lambda_I',n_I')\rangle= \delta_{0,\lambda n}$,
using~\eqref{InnProdAbel}, are not conti\-nuous in $\lambda$ at $\lambda=0$.
In the non-Abelian case, the argument is more complicated because
$\langle(\lambda_I',j_I',m_I',n_I')|\rho_{\lambda,j}(g_I)^m_n |(\lambda_I',j_I',m_I',n_I')\rangle$ need not
be zero for $\lambda\not=0$.
The value rather depends on the multiplicity of the trivial representation in the tensor product
$\rho_{j_I'}^{\otimes 2q}\otimes \rho_j^{\otimes p}$ if $\lambda_I'=q/r$ and~$\lambda=p/r$ with their least
common denominator $r$.
For $\lambda=0$, we pick out the trivial representation in $\rho_{j_I}^{\otimes 2q}$; for $\lambda\not=0$
we pick out the coef\/f\/icients of all irreducible representations of $\rho_{j}^{\otimes p}$ in~$\rho_{j_I}^{\otimes 2q}$.
In general, these coef\/f\/icients are not continuously related at~$\lambda=0$.

}

Instead of using derivatives and connection operators, we are required to use holonomies
$\rho_{\delta,j}(g_I)^m_n$ with some f\/inite $\delta$ to construct the Hamiltonian constraint operator,
appealing to the standard relationship between curvature components and holonomies around small closed
loops.
Since the same relationship is used for the full constraint~\cite{RS:Ham,QSDI}, there is at least
a~plausible connection between models and the full theory.
We have
\begin{gather}
\label{HolExp}
h_{\Delta}=1+\ell^2s^a_1s^b_2F_{ab}^i\tau_i+O\big(\ell^4\big)
\end{gather}
if the loop $\Delta$, spanned by two unit vectors $s_1^a$ and $s_2^a$, is of coordinate area $\ell^2$.
In lattice constructions, one may use loops around elementary plaquettes, although consistency issues of
the constraint algebra may require more complicated routings (see e.g.\ \cite{TwoPlusOneDef}).
The coordinate area of a~loop in the $(I,J)$-plane is then close to $\ell_I\ell_J$, and we are led to use
homogeneous holonomies $\rho_{\delta_I,j_I}(g_I)$ with $\delta_I=\ell_I/L_I=1/{\cal N}_I$.

In Bianchi models, we have the Hamiltonian constraint~\eqref{HamHom}, or~\eqref{H} when diagonalized.
Putting holonomies along square loops and suitable constructions of triad operators together as in the full
theory~\cite{QSDI}, we obtain, following~\cite{cosmoIII,HomCosmo},
\begin{gather}
\hat{H}=-\frac{1}{(8\pi)^2\gamma^3G^2\hbar\delta_1\delta_2\delta_3}\sum_{I,J,K}\epsilon^{IJK}{\rm tr}
\Bigl(\rho_{\delta_I,1/2}(g_I)\rho_{\delta_J,1/2}(g_J)\rho_{\delta_I,1/2}(g_I)^{-1}\rho_{\delta_J,1/2}
(g_J)^{-1}
\nonumber
\\
\phantom{\hat{H}=}{}
\times\left|\rho_{\delta_K,1/2}(g_K)[\rho_{\delta_K,1/2}(g_K)^{-1},\hat{V}]\right|\Bigr)+\hat{H}_{\Gamma}
\label{Ham}
\end{gather}
with matrix products of all holonomies involved.
Instead of $j=1/2$ one may use other irreducible representations,
or add dif\/ferent such contributions~\cite{Gaul,AlexAmbig,AmbigConstr}.
Also the values of $\delta_I$ are subject to choices whose implications we will discuss in more detail below.
One may choose f\/ixed values, or relate them to properties of the state acted on by the constraint operator.
In the latter case, the state dependence of regularized constraints in the full theory would be modeled.
For now, however, we will assume f\/ixed $\delta_I$ so that $\hat{H}$ preserves a~lattice subalgebra.

The term $\hat{H}_{\Gamma}$ in~\eqref{Ham} vanishes for the Bianchi~I model and incorporates
spin-connection terms for other models, as in~\cite{Spin}.
The commutator $|\rho_{\delta_K,1/2}(g_K)[\rho_{\delta_K,1/2}(g_K)^{-1},\hat{V}]|$ in the se\-cond line
quantizes the classical combination $\epsilon^{ijk}E^{[a}_iE^{b]}_j/\sqrt{|\det E|}$ which diverges for
degenerate triads, at classical singularities of collapse type\footnote{The absolute value around the
commutator is necessary because the classical analog $\{A_a^i,V\}=2\pi\gamma G{\rm sgn}(\det E)
\epsilon^{ijk}E^{[a}_iE^{b]}_j/\sqrt{|\det E|}$ carries a~sign factor.
The absolute value avoids parity violation (see~\cite{FermionHolst} for a~detailed discussion of parity).
Classically, the sign of $\det E$ changes whenever the sign of $\{A_a^i,V\}$ changes, and one could
multiply the latter with ${\rm sgn}(\det E)$ to avoid parity violation.
However, when quantized, $\rho_{\delta_K,1/2}(g_K)[\rho_{\delta_K,1/2}(g_K)^{-1},\hat{V}]$ and
$\widehat{\det E}$ do not commute in non-Abelian models.
Since the dynamics identif\/ies states annihilated by the commutator as degenerate ones corresponding to
classical singularities, we refer to its own sign instead of multiplying it with the non-commuting operator
$\widehat{{\rm sgn}\det E}$~\cite{DegFull}.}. Writing this operator on the right in $\hat{H}$, in the
ordering chosen in~\eqref{Ham} as in the full theory~\cite{QSDI}, implies that singular states decouple
from the dynamics: They are automatically annihilated by the constraint.
(There is a~similar factor in~$\hat{H}_{\Gamma}$, also ordered to the right.) In this way, quantum
hyperbolicity~\cite{BSCG} is realized even if no dif\/ference equation is available as an explicit
evolution equation.

\subsubsection{Abelian models and dif\/ference equations}

Dif\/ference equations are obtained in Abelian models after transforming to the triad representation,
provided the Hamiltonian constraint f\/ixes a~lattice subalgebra.
First, writing Abelian holonomies of a~diagonal Bianchi model as $\rho_{\delta_I,1/2}(g_I)=\cos(\delta_I
c_I/2)+ 2\Lambda_I^i\tau_i\sin(\delta_I c_I/2)$, one can compute all matrix products and the trace
in~\eqref{Ham}.
Since the f\/inal result is lengthy, we def\/ine
\begin{gather*}
\hat{K}_3:=\sin(\delta_1c_1/2)\cos(\delta_1c_1/2)\sin(\delta_2c_2/2)\cos(\delta_2c_2/2)=\frac{1}{4}
\sin(\delta_1c_1)\sin(\delta_2c_2)
\end{gather*}
and cyclic permutations thereof, as well as
\begin{gather*}
\hat{I}_J=\left|2i\left(\sin(\delta_J c_J/2)\hat{V}\cos(\delta_J c_J/2)-\cos(\delta_J c_J/2)\hat{V}
\sin(\delta_J c_J/2)\right)\right|.
\end{gather*}
The Hamiltonian constraint is then
\begin{gather*}
\hat{H}=-\frac{1}{8\pi^2\gamma^3G^2\hbar\delta_1\delta_2\delta_3}\sum_{J=1}^3\hat{K}_J\hat{I}_J.
\end{gather*}
It is straightforward to compute the action of $\hat{K}_J$ and the eigenvalues of $\hat{I}_J$ on the
$(\delta_1,\delta_2,\delta_3)$-lattice subalgebra (now dropping the f\/ixed $\delta_I$ from the notation of
states):
\begin{gather*}
\hat{K}_3|n_1,n_2,n_3\rangle=-\frac{1}{16}
(|n_1+2\delta_1,n_2+2\delta_2,n_3\rangle-|n_1-2\delta_1,n_2+2\delta_2,n_3\rangle
\\
\phantom{\hat{K}_3|n_1,n_2,n_3\rangle=}{}
-|n_1+2\delta_1,n_2-2\delta_2,n_3\rangle+|n_1-2\delta_1,n_2-2\delta_2,n_3\rangle)
\end{gather*}
and cyclic permutations, and
\begin{gather*}
\hat{I}_1|n_1,n_2,n_3\rangle=|V_{n_1+\delta_1,n_2,n_3}-V_{n_1-\delta_1,n_2,n_3}||n_1,n_2,n_3\rangle
\\
\phantom{\hat{I}_1|n_1,n_2,n_3\rangle}{}
=(8\pi\gamma)^{3/2}\ell_{\rm P}^3\left|\sqrt{|n_1+\delta_1|}-\sqrt{|n_1-\delta_1|}\right|\sqrt{|n_2n_3|}
|n_1,n_2,n_3\rangle.
\end{gather*}

\looseness=-1
Since $\hat{H}|n_I\rangle$ must vanish (or equal the action of a~matter Hamiltonian $\hat{H}_{\rm matter}$
on $|n_I\rangle$), a~dif\/ference equation is obtained in the triad representation of coef\/f\/icients
$\psi_{n_1,n_2,n_3}$ in $|\psi\rangle= \sum_{n_I}\psi_{n_1,n_2,n_3} |n_I\rangle$.
Introducing $s_{n_1,n_2,n_3}:= \sqrt{|n_1n_2n_3|}\psi_{n_1,n_2,n_3}$ to shorten the
expression\footnote{Dividing by $n_1n_2n_3$ is well-def\/ined for the evolution equation because
$|n_1,n_2,n_3\rangle$ is annihilated by the constraint whenever $n_1n_2n_3=0$, as part of the property of
quantum hyperbolicity.
The coef\/f\/icients $\psi_{n_1,n_2,n_3}$ with $n_1n_2n_3=0$ decouple from the rest and can safely be
ignored.}, we have
\begin{gather}
-A_{\delta_1}(n_1)\left(s_{n_1,n_2+2\delta_2,n_3+2\delta_3}-s_{n_1,n_2-2\delta_2,n_3+2\delta_3}
-s_{n_1,n_2+2\delta_2,n_3-2\delta_3}+s_{n_1,n_2-2\delta_2,n_3-2\delta_3}\right)
\nonumber
\\
\quad{} -A_{\delta_2}(n_2)\left(s_{n_1+2\delta_1,n_2,n_3+2\delta_3}-s_{n_1-2\delta_1,n_2,n_3+2\delta_3}
-s_{n_1+2\delta_1,n_2,n_3-2\delta_3}+s_{n_1-2\delta_1,n_2,n_3-2\delta_3}\right)
\nonumber
\\
\quad{} -A_{\delta_3}(n_3)\left(s_{n_1+2\delta_1,n_2+2\delta_2,n_3}-s_{n_1-2\delta_1,n_2+2\delta_2,n_3}
-s_{n_1+2\delta_1,n_2-2\delta_2,n_3}+s_{n_1-2\delta_1,n_2-2\delta_2,n_3}\right)
\nonumber
\\
=128\pi^2\gamma^3G\ell_{\rm P}^2\delta_1\delta_2\delta_3\frac{\hat{H}_{\rm matter}(n_1,n_2,n_3)}
{V_{n_1,n_2,n_3}}s_{n_1,n_2,n_3}
\label{Evolve}
\end{gather}
with volume eigenvalues inserted.
We used partial eigenvalues of the matter Hamiltonian\footnote{For simplicity, we assume the absence of
connection couplings; see~\cite{LRSParity} for a~fermionic model in which the assumption does not hold.
No qualitative changes to the present statements occur in such a~case.}
\begin{gather*}
\hat{H}_{\rm matter}|\psi\rangle=\sum_{n_I}\left(\hat{H}_{\rm matter}(n_I)\psi_{n_1,n_2,n_3}
\right)|n_I\rangle
\end{gather*}
which may still act on a~matter-f\/ield dependence of $s_{n_1,n_2,n_3}$.
The other coef\/f\/icients refer to eigenvalues of $\hat{I}_I$, def\/ining $A_{\delta_I}(n_I):=
I_I(n_1,n_2,n_3)/V_{n_1,n_2,n_3}$ such that $A_{\delta}(n)=
|\sqrt{|n+\delta|}-\sqrt{|n-\delta|}|/\sqrt{|n|}$.
Equation~\eqref{Evolve} correctly quantizes the classical terms such as
$c_2c_3a_1=c_2c_3\sqrt{|p^2p^3/p^1|}$ in the Hamiltonian constraint~\eqref{H}, using the ordering
$|p^1|^{-1} c_2c_3 \sqrt{|p^1p^2p^3|}$.

Equations such as~\eqref{Evolve} have been derived long ago~\cite{HomCosmo}, and reproduced since then many
times, in slightly dif\/ferent forms.
We have presented the derivation here with some detail because, lacking the proper notion of homogeneous
connections, it had not been realized before that the form is valid only for a~Hamiltonian constraint
operator f\/ixing a~lattice subalgebra of the homogeneous model.
If the $\delta_I$ are not f\/ixed once and for all but depend on the state acted on (and even its graph),
or if the $\delta_I$ used in the operator are not the same as those of the lattice subalgebra, the
decomposition rule~\eqref{Mult} or its Abelian analog~\eqref{MultAbel} must be used, implying ref\/inement.
The operator~\eqref{Ham} will still be valid, but its action must be re-derived, and does not easily give
rise to a~dif\/ference equation, certainly not one of constant step-size.
We emphasize that ref\/inement is realized even if the operator~\eqref{Ham} is not modif\/ied, provided
only one applies it to states not in the lattice subalgebra f\/ixed by it.

\subsubsection{Lattice ref\/inement: a~toy model}
\label{s:RefToy}

So far, we have presented the minisuperspace quantization of Hamiltonians.
For contact with the full theory, we must try to reduce an inhomogeneous operator and face the averaging
problem.
This task, at present, cannot be done in detail, but its outcome will af\/fect the choice of $\delta_I$.
Instead of deriving these values and their relation to states, we must resort to suf\/f\/iciently general
parameterizations to model dif\/ferent possible reductions.

If the $\delta_I$ are not adapted to a~lattice or to the common denominator of all holonomies involved,
$\hat{H}$ will not f\/ix any lattice subalgebra, even if the $\delta_I$ are f\/ixed and not state-dependent.
Lattice ref\/inement then occurs by multiplying holonomies of dif\/ferent edge lengths and obeying the
decomposition rule~\eqref{Mult}.
Evaluations of the Hamiltonian constraint, especially the non-Abelian version, become more involved and
dif\/ference equations no longer are readily available, but the property of lattice ref\/inement can be
seen already in an admittedly rough toy model.
The model is certainly far from an actual derivation from some full Hamiltonian constraint.
Instead, it is meant to illustrate how the behavior of $\delta_I$ could follow from some discrete dynamics.
Although we use the language of inhomogeneous lattice states and operators, the actual dynamics is chosen
only for illustrative purposes.
Nevertheless, there are some interesting features which may be promising for a~more faithful representation
of the loop dynamics.

Let us assume that each $\delta_I$ is always half the maximum $\lambda_I$ encountered in
a~lattice-subalgebra state acted on.
In terms of an inhomogeneous lattice, this means that every new edge generated by a~vertex contribution of
the Hamiltonian constraint would go half-way to the next vertex.
In an inhomogeneous lattice, the presentation of ref\/inement depends on the order in which individual
holonomies or vertex contributions of the Hamiltonian constraint act.
If ref\/inement proceeds regularly, staying close to cubic lattices of nearly constant link lengths, one
would expect that all plaquettes will f\/irst be split half-way along edges, and when this has happened for
all of them, one would proceed to quarters and so on.
However, in the Hamiltonian constraint all vertex contributions appear in superposition, not in
simultaneous action on a~single lattice.
To realize an ordering, one may assume that a~physical state annihilated by the constraint is expanded in
spin-lattice states according to the eigenspace of some operator such as the total volume, by the maximum
spin on all edges, or by the number of plaquettes.
Ordering spin-lattice contributions in a~physical state with respect to any of these values, plaquettes
will be f\/illed in a~certain arrangement, such as the one described.

\looseness=-1
Back in our homogeneous model, starting with a~state in some lattice subalgebra with va\-lues~$\lambda_I^{(0)}$, the f\/irst action of the Hamiltonian constraint, multiplying with holonomies of lengths~$\lambda_I^{(0)}/2$ in dif\/ferent directions, requires a~decomposition~\eqref{Mult} of the whole state,
ref\/ining edge lengths to halves.
After a~single multiplication, no homogeneous holonomy of the original length $\lambda_I^{(0)}$ will occur
explicitly, but such edges are still present in a~corresponding homogeneous lattice because there are
several matrix products $\sum_k \rho_{\lambda_I^{(0)}/2,j}(g_I)^m_k
\rho_{\lambda_I^{(0)}/2,j}(g_I)^k_n$ of two $\lambda_I^{(0)}/2$-holonomies without intermediate factors.
Taking these holonomies into account, we keep acting with $\lambda_I^{(0)}/2$-holonomies until all those
products disappear.
(In inhomogeneous lattice language, we f\/ill all $\lambda_I^{(0)}$-plaquettes with new edges and vertices
at all midpoints of the original edges.) Once these options have been exhausted, the next ref\/inement step
is due, going to $\lambda_I^{(0)}/4$ until all the previously ref\/ined plaquettes have been f\/illed.
In the process just described, we have assumed a~certain ordering of the actions of individual vertex
contributions, f\/irst f\/illing all the $\lambda_I^{(0)}$-plaquettes, then moving to
$\lambda_I^{(0)}/2$-plaquettes, and so on, as we would do on an inhomogeneous lattice.

\looseness=-1
We can relate the number of plaquettes, or the degree of ref\/inement, to geometrical quantities of the
whole lattice.
Starting with a~nearly homogeneous lattice with all edge spins $j_0$ equal, the initial area in the
$J,K$-plane is approximately $A_0=(\lambda_I^{(0)}j_0)^2 {\cal N}_J{\cal N}_K$, with ${\cal N}_J{\cal N}_K$
plaquettes in this plane and transversal links of the size $\lambda_I^{(0)}= L_I/{\cal N}_I$.
Thus, $A_0=L_I^2 j_0^2 {\cal N}_J{\cal N}_K/{\cal N}_I^2$, or $A_0\approx V_0^{2/3}j_0^2$ for a~nearly
isotropic lattice with equal edge numbers in the three directions.
When all these plaquettes have been ref\/ined after the f\/irst stage, the maximum spins have changed to
$j_1=2j_0+1/2$: multiplying $j_0$ with two because we decompose holonomies halfway according
to~\eqref{Mult}, doubling them over, and adding $1/2$ from the action of a~new holonomy in the fundamental
representation.
The added $1/2$ will soon be irrelevant when $j$ becomes larger by repeated doubling.
The area has then increased to $A_1=(\lambda_I^{(1)})^2(2j_0)^2 (2{\cal N}_J)(2{\cal N}_K)$ with
$\lambda_I^{(1)}= \lambda_I^{(0)}/2$.
Combining these equations, $A_1=4A_0$, in which only the increased spin due to ref\/inement contributes.
After $N$ steps, the same arguments show that the area has increased to $A_N=2^N A_0$, and
$\lambda_I^{(N)}= \lambda_I^{(0)}/2^N= \lambda_I^{(0)} \sqrt{A_0/A_N}$.
The spin quantum numbers increase by $j_N\approx 2^Nj_0$.

\looseness=1
It is an interesting feature that the spin of a~single action (here $1/2$), an ambiguity parameter of the
full constraint, becomes progressively less important in the model as ref\/inement proceeds, increasing
$j_N$.
The large-scale behavior is insensitive to details of the microscopic dynamics and associated ambiguities,
a~property that makes ef\/fective and mean-f\/ield viewpoints meaningful.

In this model, the edge lengths $\lambda_I^{(N)}$ are inversely proportional to the square root of the
area, or to a~linear measure of the extension of the lattice.
With near isotropy, this scaling, $\lambda=\lambda_0/a$, is of advantage for holonomy-modif\/ied dynamics,
in which holonomies $\exp(i\lambda c)=\exp(i\lambda_0 c/a)$ depend on the isotropic connection component
$c$ only in the combination $c/a\propto {\cal H}$ proportional to the Hubble parameter; the same behavior
has been proposed in~\cite{APSII} as an ad-hoc choice for the actual dynamics of loop quantum cosmology.
While $c$ may grow large even at small curvature, for instance if there is a~positive cosmological
constant, ${\cal H}$ remains small in low-curvature regimes.
The ref\/ined dynamics, with a~non-constant $\lambda$, is more well-behaved in semiclassical regimes.

In diagonal anisotropic models, the ref\/inement behavior described here implies that holono\-mies depend on
the connection by the combinations $c_I/\sqrt{|p^I|}$, as in~\cite{BianchiIBounce}, $|p^I|$ being
proportional to the area $A$ of the plane transversal to the $I$-direction.
This ref\/inement is problematic in terms of stability properties of the dif\/ference equation it
implies~\cite{SchwarzN}.
A dif\/ferent ref\/inement scheme in which $\lambda_I$ is inversely proportional to the length of the
$I$-direction is preferable~\cite{ImpBianchiI,SchwarzN, SchwarzNHol}; a~more advanced action of the
Hamiltonian constraint not f\/ixing a~lattice subalgebra would be required, a~ref\/inement scheme in which
$\delta_I$ depends not only on the length $\lambda_I$ of its own direction but also on the other two links
meeting at a~vertex.

\subsubsection{Dif\/ference equations with mean-f\/ield ref\/inement}

Lattice ref\/inement is the homogeneous realization of discrete dynamical processes in the full theory;
ideally, its form would be derived by reducing a~full Hamiltonian constraint.
A dynamical state of quantum gravity should in general be expected to have dif\/ferent lattice structures
and spacings at dif\/ferent times, or on dif\/ferent spatial slices, especially in loop quantum gravity
whose Hamiltonians are generically graph-changing.
The number of lattice sites is then a~dynamical parameter.
Indeed, if the ${\cal N}_I$ or the $\ell_I$ are kept constant~-- we always assume $L_1$, $L_2$ and $L_3$
to be constant as these are classical auxiliary parameters~-- cosmic expansion would quickly blow up the
discreteness scale, $\ell_I\sqrt{|p^1p^2p^3|}/|p^1|$ as measured in a~diagonal Bianchi geometry, to
macroscopic sizes.
Lattice ref\/inement must be a~key feature of quantum-gravity dynamics.
Dynamical minisuperspace operators such as the Hamiltonian constraint should not refer to constant ${\cal
N}_I$ or $\delta_I$, as assumed so far in~\eqref{Ham} and~\eqref{Evolve}, but to parameters that depend on
the total volume or the scale factor via an evolving discrete state.

Strict dif\/ference equations of loop quantum cosmology then do not exist, even in Abelian models, and
approximations cannot always be derived easily.
The correct evolution equation in a~triad representation would rather have to implement the changing number
of degrees of freedom, a~problem studied in other contexts as well~\cite{CanSimp,EvolvingHilbert,PH}.
Instead of working with such complicated equations, there are two approximation schemes that help to f\/ind
properties of solutions: Ef\/fective equations and dif\/ference equations in redef\/ined variables.
\begin{description}\itemsep=0pt
\item[Ef\/fective equations] describe properties of solutions of dif\/ference equations in Abelian mo\-dels
via a~non-canonical basic algebra, the discreteness implemented by using exponentials of the connection.
For instance, we would represent a~discrete degree of freedom $(c,p)$ by a~non-canonical basic pair
$(\exp(i\delta c),p)$ with a~closed linear algebra under Poisson brac\-kets.
If $\delta$ depends on $p$ by a~power law $\delta(p)=\delta_0 |p|^x$ as a~form of lattice ref\/inement,
$(\exp(i\delta_0 |p|^xc),|p|^{1-x})$ still satisf\/ies a~closed linear algebra~\cite{BounceCohStates}.
We then generate evolution by a~Hamiltonian much like~\eqref{Ham}, depending on $\exp(i\delta(p) c)$
according to the regularization chosen.
Ef\/fective quantum evolution equations then follow the general scheme of~\cite{EffCons, EffAc} and provide
approximate information about ref\/ining solutions.
While strict dif\/ference equations are not available in lattice-ref\/ining Abelian or in non-Abelian
models, ef\/fective equations can still be formulated and solved in both cases.
\item[Approximate dif\/ference equations] in re(de)f\/ined variables model ref\/ined quantum evo\-lu\-tion by
dif\/ference equations equally spaced not in the original triad eigenvalues, but rather in some redef\/ined
versions obtained as non-linear functions of them, such as power laws.
One can derive a~suitable equidistant parameter if one knows how the $\delta_I$ depend on $n_J$ in the
ref\/ining case.
Instead of an equation~\eqref{Evolve} with $n_J$-dependent increments, for instance in
$s_{n_1,n_2+2\delta_2(n_1,n_2,n_3), n_3+2\delta_3(n_1,n_2,n_3)}$, one can sometimes work with an
equidistant dif\/ference equation in re(de)f\/ined independent variables.
If $\delta_I$ depends only on $n_I$ with the same value of $I$, we def\/ine $\bar{n}_I(n_I):=\int_0^{n_I}
(\delta_I(z))^{-1}{\rm d}z$ such that $\bar{n}_I(n_I+\delta_I(n_I))= \int_0^{n_I+\delta_I(n_I)}
(\delta_I(z))^{-1}{\rm d}z= \bar{n}_I(n_I)+ \int_{n_I}^{n_I+\delta_I(n_I)} (\delta_I(z))^{-1}{\rm d}z=
\bar{n}_I(n_I)+ \delta_I(n_I)(\delta_I(n_I))^{-1}(1+O(\delta_I'(n_I)))=
\bar{n}_I(n_I)+1+O(\delta_I'(n_I))$, a~constant increment in regions in which the derivative
$\delta_I'(n_I)$ is suf\/f\/iciently small.
If $\delta_I(n_I)\propto |n_I|^x$ is a~power law with $x<0$ for ref\/inement, the equidistant approximation
is good at large~$n_I$ but not for small~$n_I$, where the quantum dynamics remains ambiguous, anyway.
(For $\delta_I\propto|n_I|^x$, we have an equidistant equation in $\bar{n}_I\propto |n_I|^{1-x}$,
corresponding to the new $p$-dependent variable used for ef\/fective equations.) One may also redef\/ine
the whole dif\/ference equation in terms of $\bar{n}_I$ with constant increments, dropping
$O(\delta_I'(n_I))$-terms as a~specif\/ic choice of factor ordering~\cite{SchwarzN}.
If $\delta_I$ depends not just on $n_I$ with the same $I$, a~redef\/inition is more complicated to derive.
If $\delta_1\delta_2\delta_3$ is proportional to a~power of $|n_1n_2n_3|$, $\delta_1\delta_2\delta_3\propto
|n_1n_2n_3|^x$ such that ref\/inement does not introduce additional anisotropy, one can always f\/ind one
equidistant variable given by
\begin{gather*}
N(n_1,n_2,n_3):=\int_0^{n_1}\int_0^{n_2}\int_0^{n_3}
(\delta_1(z_1,z_2,z_3)\delta_2(z_1,z_2,z_3)\delta_3(z_1,z_2,z_3))^{-1}{\rm d}z_1{\rm d}z_2{\rm d}z_3
\\
\hphantom{N(n_1,n_2,n_3)}{} \propto
|n_1n_2n_3|^{1-x}
\end{gather*}
related to the total volume~\cite{SchwarzN}.
This choice resembles Misner variables~\cite{Mixmaster}, which refer to the volume (or scale factor) and
two anisotropy parameters.
\end{description}

A state dependence of dynamical operators, underlying lattice ref\/inement, may seem unexpected.
After all, the dynamical operators are used to derive evolving states; how can properties of such states
enter the def\/inition of dynamical operators or the dif\/ference equations they imply? Taking reduction
seriously, it turns out that state dependence is unavoidable.
In minisuperspace models, we cannot formulate a~dynamical operator from f\/irst principles, or if we do so,
the results are fraught with minisuperspace artefacts because full properties of the discreteness are
ignored.
As described before, reduced dynamics is supposed to model the full dynamics of a~symmetric state, to be
projected back to the space of symmetric states after each application of the evolution operator, a~process
that includes the decomposition rule~\eqref{Mult} used crucially in our toy model of ref\/inement.
A minisuperspace evolution operator obtained by reduction must encode both the full Hamiltonian constraint
and properties of the projection.
The latter depends on the evolving state to be projected back on the symmetric space.
While the precise form remains complicated to determine, we see how a~state dependence of the end result is
obtained.

Without a~detailed method to perform dynamical reduction, the phase-space dependence of parameters such as
$\delta_I$ or ${\cal N}_I$ is inserted in equations only after the operator has been formulated, as a~kind
of mean f\/ield describing microscopic properties not directly accessible at the minisuperspace level.
Such details and ambiguities are relevant at small scales and at higher curvature, or in strong quantum
regimes.
These regimes can be understood only by general ef\/fects, such as quantum hyperbolicity, but away from
deep quantum regimes, ef\/fective and mean-f\/ield pictures are meaningful and useful.

\subsubsection{Ad-hoc modif\/ications in pure minisuperspace models}
\label{s:adhoc}

We have presented a~quantization of homogeneous models which has a~tight link with the full theory and,
unlike previously existing versions, applies in non-Abelian cases.
Lattice ref\/inement naturally arises as a~consequence of state-dependent regularizations as in the full
theory, combined with a~reduction of all states, including physical ones, to the space of homogeneous loop
quantum cosmology.

Lattice ref\/inement is important for consistent dynamics, for a~f\/ixed lattice expanded by cosmic
evolution would either be coarse at the present time, or would have to start at tiny spacings, orders of
magnitude below the Planck length, to be unnoticeable in current observations.
Lattice ref\/inement as a~dynamical process ensures that the discreteness scale does not need to follow
cosmic expansion; it can remain small at a~constant or slowly-changing value as macroscopic events happen
on larger regions.
With a~concrete realization of lattice ref\/inement, we can look back at minisuperspace modif\/ications
that have been proposed in the hope of obtaining appropriate dynamics, and see how justif\/ied their
assumptions are from the perspective of the new picture.
The most commonly used ad-hoc modif\/ication is a~change of classical basic variables before isotropic
minisuperspace quantization, in which $c/a\propto \dot{a}/a$ appears in holonomies, and the role of $p$ is
played by the volume.

We f\/irst note that f\/luxes necessarily result as reduced operators in the derived basic algebra, not
other powers of densitized-triad components or the volume\footnote{It is possible to construct f\/lux
operators from the volume operator, viewing the latter as some kind of basic operator~\cite{Flux}.
However, for the present purposes one cannot substitute the volume for f\/luxes because no linear basic
algebra would result for the def\/inition of quantum representations and their averaging and reduction.}.
Basic operators or linear functions of them are directly reduced by reference to the commutation result of
Section~\ref{s:comm}.
Non-linear functions such as the volume, on the other hand, are more complicated to average or reduce
exactly, with no currently known procedure to do so\footnote{Moreover, the volume operator usually used in
homogeneous models, and also here, is a~simplif\/ied version of the cubic SU(2)-invariant of the full
theory.
The assumed simplif\/ication of the much more complicated full spectrum does not follow from reduction but
is put in by hand.
When details of the eigenvalues are important, for instance when one uses the volume as the independent
variable of dif\/ference equations, the simplif\/ied spectrum could lead to additional artefacts, not
covered by the methods of this article.}. The basic holonomy operators therefore act by shifts on the f\/lux
spectrum by constant amounts (of $p$ in isotropic models), not the volume spectrum.
If constant shifts of the volume spectrum have dynamical advantages, as in the model of~\cite{APSII}, they
cannot be derived by direct use of basic operators but only after re(de)f\/ining variables as in the
preceding subsection.
The volume can be used as a~basic variable only as a~modif\/ication within a~pure minisuperspace
quantization, without reduction and a~justif\/ied analog in the full theory.

\looseness=-1
The modif\/ications proposed in~\cite{APSII} have been motivated in holonomy-based expressions for
$F_{ab}^i$ in the constraint by referring to geometrical areas $a^2\ell^2$ instead of coordinate ones
$\ell^2$, where $a$ is the scale factor of a~Friedmann--Lema\^{\i}tre--Robertson--Walker model to be
quantized.
However, just as tensor components $F_{ab}^i$ depend on coordinates, it is the coordinate area $\ell^2$
which should be used in the expansion~\eqref{HolExp}, not geometrical areas obtained using the metric or
densitized triad.
(Contracting $F_{ab}^i$ with the two vector f\/ields provides a~scalar.
However, for the coordinate area $\ell^2$ to be the correct factor in the expansion, the vector f\/ields
must be normalized using a~background metric.
Changing coordinates and retaining normalization then makes the contracted version transform.)

If these and other ad-hoc assumptions, for instance about factor ordering, are dropped, dynamical equations
are much more ambiguous than usually realized or admitted.
More-involved constructions of lattice ref\/inement are required, which capture necessary projections of
the dynamical f\/low back on the space of symmetric states.
Exact projections being largely unknown, the dynamics can be obtained only in parameterized ways,
faithfully taking into account ambiguities~\cite{InhomLattice,CosConst}.
At this dynamical stage, the construction of minisuperspace operators currently proceeds by analogy with
full operators, not by derivation.

\subsection{Quantum-geometry corrections}

Using holonomies instead of curvature or connection components implies quantum-geometry corrections in the
dynamics.
There is a~second type of ef\/fect, called inverse-triad correction, which comes from the fact that an
inverse of the densitized triad appears in the Hamiltonian constraint of gravity and in matter
Hamiltonians, but f\/lux operators have discrete spectra containing zero.
No inverse f\/lux operators exist, but the inverse densitized triad can be quantized to a~densely def\/ined
operator using classical rewritings following~\cite{QSDI,QSDV}.
In the Hamiltonian constraint, inverse-triad operators appear in the gravitational part (giving rise to
dif\/ferences of volume eigenvalues in~\eqref{Evolve} and in matter Hamiltonians).

Holonomy corrections are controlled by the parameters $\lambda_I$, or by the values $\delta_I$ chosen for
a~constraint operator.
Inverse-triad operators entering~\eqref{Ham} via $|\rho_{\delta_K,1/2}(g_K)
[\rho_{\delta_K,1/2}(g_K)^{-1},\hat{V}]|$ are built using the same type of holonomies, and so their
corrections refer to the same lattice scales $\delta_I$ as holonomy corrections.
Both corrections are therefore linked to each other, and comparing the explicit forms of corrections allows
one to estimate which one might be dominant in a~given regime.

Like holonomy corrections, the size of inverse-triad corrections depends on the values of $\delta_I$ and
requires a~proper consideration of lattice structures.
However, there is an additional operator, the volume $\hat{V}$, used crucially in the def\/inition of
inverse-triad operators as commutators.
For this operator, the same question must be asked as for holonomies, namely what lattice scale it refers
to.
In a~local lattice picture, as in the full theory, one should expect the relevant volume to be the one
associated with a~single lattice site or a~spin-lattice vertex, just as the holonomies used correspond to
single lattice links.
However, incorporating the volume in this way is not as obvious as for holonomies, and so dif\/ferent
versions have been considered, making use of macroscopic volumes~\cite{ScalarHolEv} or even one associated
with the artif\/icial integration region ${\cal V}$.
In this subsection, we derive in detail the form of inverse-triad operators and the corrections they imply.
To simplify commutator calculations involved, we will present the main equations for Abelian models and
brief\/ly comment on non-Abelian ef\/fects later.

\subsubsection{Local and non-local lattice operators}

Working with lattice spin-network states, one can def\/ine dif\/ferent f\/lux operators which all give rise
to the same f\/lux when averaged to minisuperspace operators.
This situation complicates constructions in pure minisuperspace models and has led to considerable
confusion.
Only relating models to the full theory, completing the kinematical reduction, can solve these issues.
\begin{description}\itemsep=0pt
\item[Local lattice operators:] We begin with the local f\/lux operator, able to show any inhomo\-genei\-ty
realized in the lattice model: $\hat{F}_{v,I}$, taken for a~plaquette transversal to a~sur\-fa\-ce~$X^a_I$ and
intersecting only one edge $e_{v,I}$ starting at the vertex~$v$.
We choose the surface to be a~square of coordinate area $\ell_J\ell_K=\lambda_J\lambda_KL_JL_K$, so that we
can view $\hat{F}_{v,I}$ as a~quantization of the classical $\lambda_J\lambda_Kp^I(v)$, where $p^I(v)$ is
an inhomogeneous diagonal component making the homogeneous variables position dependent.
The conjugate variable $\lambda_Ic_I$ is quantized via local holonomies $h_{v,I}=\exp(i\int_{e_I} c_I{\rm
d}s)$.
(Recall our Abelian simplif\/ication in this subsection.) These local lattice operators satisfy the
commutator algebra
\begin{gather}
\label{Local}
[\hat{h}_{v,I},\hat{F}_{v',I'}]=-8\pi\gamma\ell_{\rm P}^2\delta_{v,v'}\delta_{I,I'}\hat{h}_{v,I}.
\end{gather}
\item[Minisuperspace operators:] If each surface used for local f\/lux operators is centered at the
intersection point with $e_{v,I}$, the union of all those that have the same $I$-coordinate as $v$ form
a~surface stretching through the whole integration region ${\cal V}$, without overlap of non-zero measure.
Including an average in the transversal direction, we can view the lattice sum
$\hat{\overline{p}}{}^I={\cal N}_I^{-1}\sum_v \hat{F}_{v,I}$ according to~\eqref{DerivAv} as the
f\/lux quantizing the minisuperspace variable $p^I=L_JL_K\tilde{p}^I$.
Its conjugate variable $c_I=L_I\tilde{c}_I$ in minisuperspace is quantized by holonomies, $h_I=\exp(i c_I)$
for an edge stretching through the whole integration region in direction $X^I_a$.
We have the commutator
\begin{gather}
\label{MiniAlg}
[\hat{h}_I,\hat{p}^J]=-8\pi\gamma\ell_{\rm P}^2\delta_I^J\hat{h}_I
\end{gather}
for minisuperspace operators, correctly quantizing~\eqref{PoissonAbel}.
\item[Non-local operators:] There is a~version of operators between local lattice and minisuperspace ones.
We can average local lattice f\/luxes $\hat{F}_{v,I}$ over the lattice rather than sum as in
$\hat{\overline{p}}{}^I$, or reduce the size of the minisuperspace f\/lux $\hat{\overline{p}}{}^I$ by
dividing by the number of vertices in a~surface, and def\/ine
\begin{gather}
\label{Mini}
\widehat{\overline{F_I}}=\frac{1}{{\cal N}_1{\cal N}_2{\cal N}_3}\sum_{v}\hat{F}_{v,I}=\frac{1}{{\cal N}
_J{\cal N}_K}\hat{\overline{p}}{}^I.
\end{gather}
This f\/lux operator refers to the lattice spacing but, via averaging, includes all lattice sites in the
integration region.
We will call it the non-local f\/lux operator.
With a~local holonomy, it obeys the commutator relation
\begin{gather}
\label{LocalNonLocal}
[\hat{h}_{v,I},\widehat{\overline{F_J}}]=-\frac{8\pi\gamma\lP^2}{{\cal N}_1{\cal N}_2{\cal N}_3}\delta_{IJ}
\hat{h}_{v,I}.
\end{gather}

In~\eqref{LocalNonLocal}, the number ${\cal N}_1{\cal N}_2{\cal N}_3$ of lattice sites in a~region ${\cal
V}$ replaces the coordinate vo\-lu\-me~$V_0$ of~\eqref{PoissonAbel}.
At a~technical level, $1/{\cal N}_1{\cal N}_2{\cal N}_3$ comes about as the product of $1/{\cal N}_J{\cal
N}_K$ in the plane average~\eqref{Mini}, and a~factor of $1/{\cal N}_I$ because only one out of ${\cal
N}_I$ lattice links along direction $I$ provides a~non-zero commutator
$[\hat{h}_{v,I},\hat{F}_{v',I}]\propto \delta_{v,v'}$ according to~\eqref{Local}.
\end{description}

It may seem questionable to use local holonomies and non-local f\/luxes within the same
setting~\eqref{LocalNonLocal}, but a~consistent and closed algebra of basic operators is obtained in this
way (provided the ${\cal N}_I$ are f\/ixed).
Whether such operators are meaningful physically is another question which we will soon discuss.
For now, our motivation for looking at such a~mix of local and non-local operators is that it has been used
(implicitly or explicitly) in several proposals to formulate inverse-triad corrections.

{\sloppy Properties of basic operators in the dif\/ferent algebras can be translated into one another and are
mutually consistent.
If $|(\mu_I)\rangle$ denotes non-local f\/lux eigenstates with $\widehat{\overline{F_J}}|(\mu_I)\rangle=
8\pi\gamma\lP^2 \mu_J|(\mu_I)\rangle$, the holonomy-f\/lux algebra~\eqref{LocalNonLocal} determines the
action $\hat{h}_{v,I}|\mu_I\rangle= |(\mu_I+1/{\cal N}_1{\cal N}_2{\cal N}_3)\rangle$ of local holonomies.
Constant shifts of f\/lux eigenvalues result for a~f\/ixed lattice.
The quantized densitized-triad component $p^I=L_JL_K\tilde{p}^I$ is obtained from the averaged f\/lux as
$\hat{\overline{p}}{}^I={\cal N}_J{\cal N}_K \widehat{\overline{F_I}}$.
Its eigenvalues change under the action of a~basic holonomy operator by $p^I= 8\pi\gamma\lP^2 {\cal
N}_J{\cal N}_K \mu_I \mapsto 8\pi\gamma\lP^2 {\cal N}_J{\cal N}_K (\mu_I+ 1/{\cal N}_1{\cal N}_2{\cal
N}_3)= p^I+1/{\cal N}_I$.
This dependence of the constant shift on ${\cal N}_I$ is consistent with the form $\exp(i c_I/{\cal N}_I)$
of local holonomies $\exp(i\ell_I\tilde{c}_I)$, with $c_I=L_I \tilde{c}_I$, $\ell_I/L_I= 1/{\cal N}_I$, and
$\{c_I,p^J\}= 8\pi\gamma G \delta_I^J$.
Notice the dif\/ferent behaviors of the non-local averaged f\/lux $\widehat{\overline{F_I}}$ and the
minisuperspace densitized-triad component $\hat{\overline{p}}{}^I$, corresponding by its factor of $L_JL_K$
to the f\/lux through a~complete plane in the region~${\cal V}$.

}

The general form of the algebra of basic operators does not depend much on whether the local~$\hat{F}_I$,
the non-local~$\widehat{\overline{F_I}}$ or the minisuperspace $\hat{p}^I$ is used; the latter two dif\/fer
from each other just by constant factors at the kinematical level (disregarding lattice ref\/inement).
However, $\widehat{\overline{F_I}}$ and~$\hat{p}^I$ are much less local then~$\hat{F}_{v,I}$ and therefore
unsuitable for local expressions such as quantized Hamiltonians or inverse-triad corrections.

\subsubsection{Inverse-triad corrections}

The gravitational part of the Hamiltonian constraint contains a~factor of $\epsilon^{ijk}\epsilon_{abc}
E^b_jE^c_k/\sqrt{|\det E|}$ in which one divides by the determinant of $E^a_i$, and similar terms occur in
matter Hamiltonians.
Flux operators and the volume operator having zero in their discrete spectra, no densely def\/ined inverse
exists to quantize $1/\det E$ directly.
Instead, one makes use of the classical identity~\cite{QSDI}
\begin{gather}
\label{InvClass}
2\pi\gamma G{\rm sgn}(\det E)\frac{\epsilon^{ijk}\epsilon_{abc}E^b_jE^c_k}{\sqrt{|\det E|}}
=\left\{A_a^i,\int\sqrt{|\det E|}{\rm d}^3x\right\}
\end{gather}
and quantizes the Poisson bracket to a~commutator of the form
\begin{gather}
\label{InvQuant}
\frac{1}{i\hbar}\left(\hat{h}_e^{-1}[\hat{h}_e,\hat{V}]-\hat{h}_e[\hat{h}_e^{-1},\hat{V}]\right)=\frac{1}
{i\hbar}\left(\hat{h}_e\hat{V}\hat{h}_e^{-1}-\hat{h}_e^{-1}\hat{V}\hat{h}_e\right).
\end{gather}

In a~lattice model, holonomies refer to lattice links, or $\rho_{\delta,j}(g_I)$ in their reduction.
The vo\-lu\-me operator is expressed via f\/luxes, and here local f\/lux operators are used, given the local
form of the classical Poisson bracket in~\eqref{InvClass} and of the commutator in~\eqref{InvQuant} which
depends only on vertex contributions to $\hat{V}$ lying on the edges used in $\hat{h}_e$.
At this stage, minisuperspace models can easily become misleading because their most immediate f\/lux
operators~$\hat{p}^I$ or~$\hat{\overline{p}}{}^I$, proportional to $\hat{\overline{F_I}}$ are non-local.
The wrong form and size of inverse-triad ef\/fects then results.

We now present the detailed derivation of inverse-triad corrections based on local f\/lux operators, as
in~\cite{Springer,InflTest}, and then show how non-local versions dif\/fer.
The simplif\/ied volume operator of Abelian models is $\hat{V}=\sum_v
|\hat{F}_{v,1}\hat{F}_{v,2}\hat{F}_{v,3}|^{1/2}$, summed over all vertices of a~spin-lattice state.
In expressions such as~\eqref{InvQuant}, it suf\/f\/ices to look at contributions from all lattice-aligned
$\hat{h}_{v,I}$.
A~single such commutator is then
\begin{gather*}
\hat{I}_{v,I}
=\frac{\big|\hat{h}_{v,I}^{\dagger}\hat{V}\hat{h}_{v,I}-\hat{h}_{v,I}\hat{V}\hat{h}_{v,I}^{\dagger}\big|}
{8\pi\gamma G\ell_{\rm P}^2}=\frac{\left|\hat{h}_{v,I}^{\dagger}\sqrt{|\hat{F}_{v,I}|}\hat{h}_{v,I}
-\hat{h}_{v,I}\sqrt{|\hat{F}_{v,I}|}\hat{h}_{v,I}^{\dagger}\right|}{8\pi\gamma G\ell_{\rm P}^2}
\sqrt{|\hat{F}_{v,J}\hat{F}_{v,K}|}
\end{gather*}
to \looseness=1 be summed over all $I$.
(Classically, the combination of holonomies and f\/luxes corresponds to $|F_{v,I}|^{-1/2}
\sqrt{|F_{v,J}F_{v,K}|}$.
We have used an absolute value around the commutator as in~\eqref{Ham}.) For Abelian holonomies it is easy
to simplify the inverse-triad operator, making use of the commutator $[\hat{h}_{v,I}, \hat{F}_{v,I}]=
-8\pi\gamma\ell_{\rm P}^2 \hat{h}_{v,I}$ from~\eqref{Local} and the reality condition
$\hat{h}_{v,I}^{\dagger}\hat{h}_{v,I}=1$.
Commuting holonomies past f\/lux operators then gives $\hat{h}_{v,I}^{\dagger}|\hat{F}_{v,I}|^{1/2}
\hat{h}_{v,I}= |\hat{F}_{v,I}+8\pi\gamma\ell_{\rm P}^2|^{1/2}$, and therefore
\begin{gather}
\label{Inv}
\hat{I}_{v,I}=\frac{\left|\sqrt{|\hat{F}_{v,I}+8\pi\gamma\ell_{\rm P}^2|}-\sqrt{|\hat{F}_{v,I}
-8\pi\gamma\ell_{\rm P}^2|}\right|}{8\pi\gamma G\ell_{\rm P}^2}\sqrt{|\hat{F}_{v,J}\hat{F}_{v,K}|}.
\end{gather}

In strong quantum regimes, non-Abelian features should be relevant~\cite{BoundFull} and inverse-triad
ef\/fect compete with holonomy and higher-curvature terms; however, the form~\eqref{Inv} still plays
a~characteristic role in ef\/fective actions~\cite{Action}.
The expression~\eqref{Inv} is a~good approximation in perturbative settings with
$F_{v,I}>8\pi\gamma\ell_{\rm P}^2$, where it may be used to estimate
qualitative ef\/fects or potential observational tests~\cite{LoopMuk,InflConsist,InflTest}.

Since inverse-triad operators are local~-- commutators $\hat{h}_e[\hat{h}_e^{-1},\hat{V}]$ provide
contributions only for vertices on $e$ even if the volume operator for the full region ${\cal V}$ is
used~-- their commutators refer to local $\hat{F}_{v,I}$ in $\hat{V}=\sum_v
|\hat{F}_{v,1}\hat{F}_{v,2}\hat{F}_{v,3}|^{1/2}$, not to the minisuperspace operator $\hat{p}^I$ or the
non-local $\widehat{\overline{F_I}}$.
Inverse-triad corrections therefore depend on $F_{v,I}\pm 8\pi\gamma\lP^2$, where the Planckian addition
can easily be a~signif\/icant contribution to the eigenvalue or expectation value of $\hat{F}_{v,I}$, the
f\/lux through an elementary lattice site\footnote{In fact, if $F_{v,I}$ is Planckian, with lattice spins
near $1/2$ for the fundamental representation, as often assumed, inverse-triad corrections are large.
Geometry must be suf\/f\/iciently excited above fundamental spins (some kind of ground state) for good
semiclassical states to result.}. Had we used the average $\widehat{\overline{F_I}}$, the algebra would have
led us to $\overline{F_I}\pm 8\pi\gamma\lP^2/{\cal N}_I{\cal N}_J{\cal N}_K$, with corrections not only
much suppressed by dividing by the large number of lattice sites but also depending on the size of the
arbitrary region ${\cal V}$ chosen\footnote{Sometimes, it is suggested to take a~limit of $V_0\to\infty$,
or ${\cal N}_I\to\infty$, viewing a~f\/inite $V_0$ as a~regulator.
The procedure removes any ${\cal V}$ dependence and makes inverse-triad corrections disappear.
However, as discussed in more detail in the next section, this reasoning is misguided: $V_0$ is not
a~regulator because its value does not at all af\/fect the classical theory.
Classical models with dif\/ferent $V_0$ produce the same physics, and so they should all be quantizable,
without an ef\/fect of $V_0$.
Moreover, the limit of ${\cal N}_I\to\infty$ is not consistent with the basic algebra of averaged
operators.}. Such operators would be incorrect; they are based on the confusion of the correct average
$\overline{\sqrt{F_I}}$ with the non-local $\sqrt{\overline{F_I}}$.

\subsubsection{Local quantum corrections}

We have distinguished three types of constructions for composite operators quantizing a~symmetric model:
the minisuperspace treatment using $\hat{h}_I=\widehat{\exp(i c_I)}$ and $\hat{p}^J$ with
algebra~\eqref{MiniAlg}, chimerical constructions with (local) link holonomies $\hat{h}_{v,I}$ but
non-local f\/luxes $\widehat{\overline{F_J}}$ with algebra~\eqref{LocalNonLocal}, and f\/inally local
lattice operators built from $\hat{h}_{v,I}$ and $\hat{F}_{v,J}$ with algebra~\eqref{Local}.

Local and minisuperspace treatments dif\/fer from each other by the order in which reduction and
composition of operators are done.
In non-local models, as in traditional minisuperspace versions, one f\/irst postulates or derives the
reduced basic operators $\hat{h}_I$ and $\widehat{\overline{F_J}}$ (or $\hat{\overline{p}}{}^J$) and their
algebra, and in a~second step constructs composite operators of the form ${\cal O}_{\text{non-local}}(\hat{h}_I,\widehat{\overline{F_J}})$ from them by simple insertions, following analogous steps
taken in the full theory.
In local quantizations, one f\/irst constructs operators ${\cal O}_{\rm
local}(\hat{h}_{v,I},\hat{F}_{w,J})$, adapting the full techniques to lattice states, and then restricts
them to a~quantized homogeneous model.
The second, local method is more complicated because it must deal with the reduction of non-basic,
composite operators or their averaging.
Tractable techniques exist only in rare cases, and therefore the main ef\/fects, for instance in the
Hamiltonian constraint, are incorporated by parameterizations or mean-f\/ield techniques as in
Section~\ref{s:RefToy}~-- an unsurprising feature given that local methods are analogous to
a~transition from microscopic Hamiltonians to tractable models of large-scale ef\/fects in condensed-matter
physics.

Despite technical dif\/f\/iculties, the local viewpoint has several clear advantages: it produces the
correct sizes of quantum corrections and naturally gives rise to lattice ref\/inement.
As already noted, the misrepresentation of quantum corrections in non-local models can easily be seen for
inverse-triad operators, or the key ingredient ${\cal O}=|F_I|^{1/2}$.
Non-local operators make use of the averaged f\/lux before taking the square root, quantizing ${\cal O}$ as
$\hat{{\cal O}}_{\text{non-local}}=|\widehat{\overline{F_I}}|^{1/2}$.
A local quantization, by contrast, leads to $\hat{{\cal O}}_{\rm local}= \overline{|\hat{F}_{v,I}|^{1/2}}$,
the over-line now indicating restriction to homogeneous states after taking the square root.
While linear combinations of the basic operators commute with averaging (Section~\ref{s:comm})~--
producing similar-looking basic algebras in local and non-local versions~-- non-linear combinations do not.
In non-linear combinations of the basic operators, drastic deviations between local and non-local operators
can therefore result, but only the local version correctly captures properties of the full theory in which
no averaging is done.

Local composite operators can be formulated only when the full lattice structure is taken into account, but
after reduction they refer to reduced degrees of freedom as suitable for a~quantization of a~classically
reduced symmetric model.
Minisuperspace and non-local operators may be formally consistent without direct reference to a~lattice,
provided one chooses the length parameter $\ell_I=\delta_IL_I$ of holonomies in some way, for instance
related to the Planck length or to full area eigenvalues, but at this stage the models become ad-hoc.
Moreover, in spite of their formal consistency, non-local composite operators do not provide the correct
form of corrections in operators that refer to f\/luxes or the volume, most importantly inverse-triad
corrections.

In addition to making inverse-triad corrections sizeable and interesting, the local treatment taking into
account inhomogeneity, has another implication regarding the physical evaluation of models.
There is just one set of parameters, $\delta_I$ or equivalently ${\cal N}_I=1/\delta_I$, which determines
the magnitude of holonomy {\em and} inverse-triad corrections.
It is not possible to ignore one of the corrections and focus only on the other, unless one can show that
a~regime of interest leads to values of $\delta_I$ that make one correction dominate the other.
In general, the two corrections are not strictly related to each other, because holonomy corrections are
sensitive to the classical curvature scale relative to the Planck density, while inverse-triad corrections
are sensitive to the local discreteness scale $F_{v,I}$ relative to the Planck area, as seen in~\eqref{Inv}.
A detailed, state-dependent analysis, taking into account inhomogeneous quantum geometry, is required to
estimate both corrections.

\section{Limitations of minisuperspace models}
\label{s:limit}

Minisuperspace models of quantum cosmology never provide exact solutions to full quantum gravity.
In some cases, deviations can be strong, for instance when unstable dynamics of neglected degrees of
freedom (a classical property) enlarges the mismatch between symmetric and less-symmetric solutions, which
at an initial time may have been the consequence only of a~mild violation of uncertainty
relations~\cite{MiniValid}.
The discreteness of loop quantum cosmology shows a~much larger class of minisuperspace limitations, for
discreteness is not easily reconciled with homogeneity.
As always, such limitations should be pointed out and discussed in detail, not to slander but to warn.

At the level of states and basic operators, homogeneous wave functions can be derived in precise terms,
using the distributional constructions of~\cite{SymmRed}, as recalled in Section~\ref{s:dist}.
Intuitively, averaging a~discrete state over a~continuous symmetry group cannot result in a~normalizable
wave function in the original Hilbert space (or even a~meaningful density matrix), but it is well-def\/ined
as a~distribution.
At this level, discreteness is not problematic and does not introduce ambiguities.
At the dynamical stage, however, discrete space-time structures with possible ref\/inement or (even if
there is no ref\/inement) reference to the local discreteness scale are more complicated and more
ambiguous, as occasionally pointed out well before loop quantum cosmo\-lo\-gy was introduced~\cite{River,UnruhTime, Weiss}.
Loop quantum cosmology has provided means to analyze such situations.

\subsection{Parameters}
\label{s:Param}

In addition to phase-space variables, a~strict minisuperspace model has only the parameter $V_0=L_1L_2L_3$
to refer to, appearing in the symplectic structure~\eqref{Poisson}.
There is no analog of $\ell_I=\lambda_IL_I= L_I/{\cal N}_I$, the discrete lattice scales.
And yet, the local quantum dynamics of the full theory, together with the quantum corrections it implies,
depend on the parameters $\lambda_I$ via holonomies around loops used to quantize $F_{ab}^i$ or along edges
used in commutators with the volume operator to quantize inverse triads.

It may not be obvious how exactly edge lengths $\lambda_IL_I$ enter quantum corrections, owing to a~certain
conceptual gap between the coordinate dependent $L_I$ or $\ell_I$ and geometrical aspects in this
background-independent formulation, as well as quantization ambiguities.
But quantum corrections certainly cannot depend on $V_0$, which is chosen at will (the coordinate size of
a~region used to reduce the symplectic structure) and knows nothing about the discrete scale.
If $V_0$ or any of its factors $L_I$ appears in quantum corrections, an artif\/icial dependence on
coordinates and the chosen region results, as well as wrong sizes of quantum ef\/fects.

The authors of~\cite{APSII} proposed to modify the strict minisuperspace dynamics in a~fashion that
successfully eliminates the $V_0$-dependence at least in holonomy corrections.
(Inverse-triad corrections could not be represented meaningfully in this scheme.) As a~consequence in
isotropic models, the Hubble parameter rather than $c=\gamma\dot{a}$ appears in holonomies, and the
discrete dynamics proceeds by constant steps of the volume $V_0a^3$, not of the densitized triad
$V_0^{2/3}p$.
Heuristically, as recalled in Section~\ref{s:adhoc}, one can argue for this scheme by identifying
geometrical areas $a^2\ell^2$, instead of coordinate areas $\ell^2$, with the Planck area when specifying
the size of holonomy modif\/ications.

Holonomy corrections depending on the Hubble parameter $\dot{a}/a$ rather than $\dot{a}$ have the advantage
of being easily coordinate independent.
If the Planck length~-- or a~parameter close to it such as the smallest non-zero area eigenvalue of the
full theory~-- is chosen as the discrete scale, modif\/ications $\sin(\lP {\cal H})/\lP$ of ${\cal H}$
result, independent of coordinates and of $V_0$.
Holonomy corrections then refer simply to the curvature radius relative to the Planck length (or, via the
Friedmann equation, the density scale relative to the Planck density), a~parameter which can easily be
estimated in regimes of interest.

As a~general scheme, however, the procedure suf\/fers from several problems:
\begin{enumerate}\itemsep=0pt
\item If the volume is used as a~basic variable, following~\cite{APSII}, one introduces a~sign choice by
hand, allowing all real values for $v=\pm V_0a^3$.
(Otherwise, $i\partial/\partial V$ is not essentially self-adjoint, and $\exp(t \partial/\partial V)$ not
unitary.
One would have to use the methods of af\/f\/ine quantum gravity~\cite{AffineQG} for acceptable
quantizations, but it has not been shown that this can be compatible with the use of holonomies.) In
f\/luxes, the sign appears automatically thanks to the orientation of triads~\cite{IsoCosmo}, and
$\exp(t\partial/\partial p)$ is unitary.
\item The Planck length in $\sin(\ell_{\rm P}{\cal H})/\ell_{\rm P}$ enters by a~mere postulate, which
cannot be avoided because the quantization, still at the minisuperspace-level, does not have access to
discrete structures.
One may use the full area spectrum to guess what the scale might be, but such a~procedure leaves open the
question of what structure or eigenvalues a~dynamical discrete state might give rise to.
Moreover, the minisuperspace area or volume spectrum does not have a~smallest non-zero eigenvalue.
The spectrum, seen in~\eqref{Deriv}, is discrete, with all eigenstates normalizable, but on the
non-separable kinematical Hilbert space the spectrum still amounts to a~continuous set of numbers as
eigenvalues.
One has to go slightly beyond minisuperspace models by referring to the full area spectrum, which does have
a~smallest non-zero eigenvalue, but still no reduction is performed.
In this way, the scheme becomes improvised, heuristic, and ad-hoc.
\item Going beyond strict minisuperspace quantizations is not a~bad thing; in fact, it is required for
realistic modeling.
However, schemes following~\cite{APSII} do not provide justif\/ications for the detailed way in which one
tries to go beyond.
Moreover, while they give rise to meaningful results for holonomy corrections, inverse-triad corrections
from~\eqref{Inv} are not modeled properly.
These corrections depend on the ratio of the discreteness scale $|\langle\hat{F}\rangle|$ to the Planck
area.
If one assumes that the discreteness scale is exactly the Planck area, inverse-triad corrections would
merely result in a~constant factor, not af\/fecting the dynamics much at f\/irst sight.
But the factor dif\/fers from one, and it has dynamical ef\/fects even if it changes just slightly.
In other attempts, $|\langle\hat{F}\rangle|$ was related to macroscopic areas~\cite{ScalarHolEv}, sometimes
even involving $V_0$, for instance by using areas related to the size of the re\-gion~${\cal V}$.
These proposals ignore the fact that there is only one discrete structure that both holonomy and
inverse-triad corrections can refer to, as well as their local nature.
\item To counter inappropriate references to $V_0$ in non-local quantizations, the parameter is sometimes
treated as a~regulator to be sent to inf\/inity after quantization.
Such a~formal limit would undo all inverse-triad corrections, leaving only the classical inverse in
dyna\-mi\-cal equations.
However, the limit does not exist at the level of operators~-- if it existed, it would result in an
inverse of the triad operator, which is not densely def\/ined.
One can perform the limit at the level of the dif\/ference equation for wave functions, or in ef\/fective
equations.
But while this is formally possible, the overall quantization procedure would no longer be coherent.
After all, Abelian loop quantum cosmology has dif\/ference equations for states because curvature is
replaced by holonomies, resulting in a~true modif\/ication $\sin(\ell_I {\cal H})/\ell_I$ for ${\cal H}$.
The limit $\ell_I\to0$ or $\lambda_I=1/{\cal N}_I\to 0$, which would send holonomy operators to derivatives
by $p$, does not exist at the operator level.
The Hamiltonian constraint is quantized to a~dif\/ference operator with non-zero step-size.
At the level of wave equations, acting with the operator on states in the triad representation, the limit
does exist and produces a~version of the Wheeler--DeWitt equation~\cite{SemiClass}.
If one insists on re\-moving the ``regulator'' $V_0$ by sending it to inf\/inity at the level of wave
equations, one should also remove the true regulators $\ell_I$ in holonomies at the same level.
The dynamics of loop quantum cosmology would then be no dif\/ferent from Wheeler--DeWitt dynamics.
In the new homogeneous quantization of this article, the limit $L_I\to\infty$ or $V_0=L_1L_2L_3\to\infty$
is impossible at f\/ixed $\delta_I$, because $g_I=\exp(L_I\tilde{\phi}(T_I))$ has no such limit.
For edge lengths $\ell_I=\delta_IL_I$ in $\rho_{\delta_I,j_I}(g_I)$ to remain f\/inite, one would have to
take the limit $\delta_I\to0$ simultaneously with $L_I\to\infty$, but then the dif\/ference equation would
become a~dif\/ferential equation, and loop quantum cosmology would, again, reduce to Wheeler--DeWitt
quantum cosmology.
Instead, one must be able to derive models for arbitrary values of $L_I$, such that observables are
independent of them.
\item The parameter $V_0$, in contrast to $\delta_I$ in holonomy modif\/ications, is not a~regulator
because it does not modify the classical theory.
Classical models can be formulated with all f\/inite choices of~$V_0$, producing the same dynamics and
observables.
(Dif\/ferent choices of~$V_0$ to some degree resemble dif\/ferent normalizations of the scale factor.
Rescaling~$V_0$ is not a~canonical transformation as it changes the symplectic structure~\eqref{Poisson}.
Classically, this is not a~problem, but one cannot expect a~unitary transformation at the quantum level,
making the issue in quantum theory more complicated.) It should then be possible to formulate also quantum
dynamics for all possible choices, or else~$V_0$ would acquire more physical meaning than it deserves.
Another problematic feature of regularization attempts is a~possible topology dependence.
If one looks at a~model of closed spatial slices, for instance the FLRW model with positive curvature, one
cannot send~$V_0$ to inf\/inity.
Instead, it may seem natural to use the full spatial coordinate volume as a~distinguished value of~$V_0$.
(But again, one may equally well formulate the classical dynamics with dif\/ferent values of~$V_0$,
choosing dif\/ferent coordinates on the unit 3-sphere or integration regions smaller than the whole
sphere.) It is sometimes argued that some ef\/fects, such as inverse-triad corrections, are meaningful or
non-zero only with closed spatial topologies, but not for the f\/lat, non-compactif\/ied FLRW model.
Not surprisingly for a~quantization based on non-local f\/luxes, quantum dynamics would then suf\/fer from
a~strong violation of locality, depending on the global spatial topology even in its elementary changes.
Such models in cosmology would also be hard to test empirically.
One would have to know the spatial volume~-- and whether it is compact or not~-- before one can estimate
quantum ef\/fects and make predictions.
A more detailed discussion is given in~\cite{Springer}.
\end{enumerate}

The many conf\/licting comments that can be found in the literature following~\cite{APSII}, for instance
regarding the size of inverse-triad corrections, attest to the complicated and incomplete state of
af\/fairs in this scheme.
Sometimes, a~single paper may claim that inverse-triad corrections are too small to be signif\/icant, and
at the same time can be changed at will by tuning the value of $V_0$.
Although it is not always realized by all authors, such conf\/licting statements spell out limitations of
pure minisuperspace models.

\subsection{Parameterizations}

By introducing ``holonomies'' as functions of the Hubble parameter rather than the connection component,
one mimics an $a$-dependent $\ell_I=\lambda_IL_I\propto 1/a=1/\sqrt{|p|}$.
As $a$ or $p$ changes and the universe expands or contracts, the lattice spacing evolves.
Although there is no explicit creation of new vertices, the number ${\cal N}$ of lattice sites must change,
for a~f\/ixed $V_0$ with changing $\lambda_I$ implies an evolving ${\cal N}=1/\lambda_1\lambda_2\lambda_3$.

Deriving a~precise functional form for $\lambda_I(p^J_i)$ would require one to formulate a~correspondence
between full discrete dynamics and reduced minisuperspace dynamics, including projections of evolved states
onto the space of symmetric states.
Lacking such complicated constructions, one can use phenomenological input to restrict possible forms of
$\lambda_I(p^J_j)$, or $\lambda_I(p^J)$ in diagonal anisotropic models, for instance dif\/ferent exponents
of power-laws $\lambda(p)\propto |p|^x$ in isotropic models with a~real parameter $x$~\cite{InhomLattice}.
If $\lambda$ is constant ($x=0$, corresponding to~\cite{IsoCosmo}), the discreteness scale would be
magnif\/ied by cosmic expansion, presumably making it noticeable in observations.
Since no discreteness has been seen, $x=0$ or values close to it are ruled out.
The suggestion of~\cite{APSII} amounts to $x=-1/2$, with constant discreteness scale, and is compatible
with observations.
However, a~constant discreteness scale is not in agreement with full constraint operators changing vertex
structures and local volume values, not just the number of vertices.
On average over many individual actions of the Hamiltonian constraint and on large scales, cosmic
minisuperspace dynamics may be close to $x=-1/2$ as in the toy model presented in
Section~\ref{s:RefToy}, but this value cannot be realized precisely.

The choice of $x=-1/2$ (or its generalization to anisotropic models) is compatible with most cases of
cosmic evolution, but it has problems with black-hole models~\cite{Consistent}.
By its construction using geometrical areas of the region ${\cal V}$, the scheme relates the number of
vertices to the total volume of spatial regions.
Near the horizon in homogeneous coordinates of the Schwarzschild interior, the spatial volume shrinks,
making the number of lattice sites small.
However, the regime is supposed to be semiclassical for large black-hole mass, which is in conf\/lict with
a~small number of lattice sites, implying noticeable discreteness.
The analysis of dif\/ferent models~-- cosmological ones and those for black holes~-- shows that there
cannot be a~single universal power-law exponent for $\lambda_I(p^J)$ in all regimes.
Discrete quantum dynamics and ref\/inement behavior, just as the underlying state, depend on the regime
analyzed.
The role of coordinate choices hints at another important issue, namely how much the condition of
covariance and anomaly freedom restricts possible ref\/inement schemes.
This question remains largely unexplored owing to the complicated nature of the quantum constraint algebra,
but see~\cite{ScalarHol} for an interesting cosmological example that suggests restrictions, also pointing
at a~value near $x=-1/2$.

These phenomenological indications notwithstanding, a~demonstration that $x=-1/2$ or a~value near it is
more than an ad-hoc choice in cosmological models would require some kind of derivation from the full
theory.
For this feat, in turn, one would need to solve the problem of the semiclassical limit of unrestricted loop
quantum gravity, which remains one of the most pressing and most complicated problems of the f\/ield.

\subsection{Reduction}

Lattice ref\/inement in dif\/ference and ef\/fective equations refers to state parameters, most importantly
$\lambda_I$, depending on a~geometrical variable such as the total volume $V$.
One may view the appearance of $V$ as an internal time, on which the evolving state depends.
A possible procedure of implementing such a~dependence, as alluded to in Section~\ref{s:RefToy},
would be to write a~full state as a~superposition $\sum_V\psi_V$ of contributions $\psi_V$ belonging
to some f\/ixed volume eigenvalue $V$.
One would decompose a~dynamical state as an expansion in eigenstates of the internal-time ope\-ra\-tor, such as
the volume.
Although the procedure would be dif\/f\/icult owing to the complicated volume spectrum and the fact that
one would have to solve for a~dynamical state f\/irst, it is in line with standard treatments of internal
time.
The states appearing in the decomposition of~$\psi_V$ in the spin-network basis then show what discrete
structures are realized at a~given vo\-lu\-me~$V$, and the spacing as well as the number of vertices might
certainly change as one moves from one~$V$ to the next.

\looseness=1
After decomposing a~dynamical state as $\sum_V\psi_V$, one would still have to adapt it to
near-homogeneous geometries, that is implement the projection back to symmetric states.
Additional state-dependent parameters may arise, all to be modeled by suitable functions
$\lambda_I(p^J_j)$, the only parameters that survive with exact homogeneity.
Such functions would then be inserted in dynamical equations of reduced models, for instance in
dif\/ference equations of Abelian \mbox{models}.

Notice that one must know some features of the full evolution of a~state before def\/ining the reduced
Hamiltonian constraint, which in a~second step can be used to evolve a~reduced state.
Since the Hamiltonian constraint is one of the operators to be averaged for reduction, evolution and
reduction are not independent processes in the construction of models.
As a~consequence, reduced Hamiltonian constraints are state-dependent, even more so than the full
constraint operator with its state-dependent regularization of~\cite{QSDI}.

If for a~precise reduction we must know how to evolve a~full state, why do we not work with the full
evolved state rather than its reductions? The advantage of reduced models is that they of\/fer additional
approximation schemes, for instance in the derivation of observables or of ef\/fective equations.
However, reduced models can never provide exact predictions~-- if their predictions were exact, one would
not be dealing with a~reduced model.
It does not make much sense to derive physical quantities, for instance bounce densities, in exact terms
within minisuperspace models because the models themselves are not exact.
Only general ef\/fects, such as quantum hyperbolicity, the presence of bounces under certain conditions, or
qualitative low-curvature corrections may be meaningful predictions, but not specif\/ic values of some
parameters related to the discreteness scale.

\subsection{Spin-foam cosmology?}

Spin-foam cosmology~\cite{SFC,EffSFC, LQCStepping} attempts to enlist spin-foam techniques to address
quantum-cosmological questions by embedding a~simple structure with f\/initely many edges in a~spatial (or
space-time) manifold $\Sigma$.
Such a~map is clearly dif\/ferent from the (mini-)superspace embedding ${\cal M}\to{\cal S}$ used for
classical reductions, or a~map $\sigma\colon {\cal H}_{\rm hom}\to {\cal D}_{\rm inhom}$ of state spaces
used for quantum reductions.
Looking back at our discussion in Section~\ref{s:Reduction}, the question therefore arises what kind
of construction spin-foam cosmology can provide.

As usually emphasized in this context, spin-foam cosmology aims at a~description of quantum cosmological
space-times without making use of reductions, rather describing physics in a~full theory of quantum gravity
in which inhomogeneous modes are still present and quantum f\/luctuate.
According to the classif\/ication in Section~\ref{s:Reduction}, such a~non-reduction scheme could
only be selection or projection, but there is certainly no control over the full non-symmetric solution
space, let alone the averaging problem, within spin-foam cosmology.
For this reason, spin-foam cosmology does not fall within our classif\/ication of reduction schemes.
If one wanted to change this conclusion, one would f\/irst have to clarify the precise relation between an
inhomogeneous amplitude and the proposed isotropic one in spin-foam cosmology.
So far, one just obtains a~new model from a~full one by inserting isotropic labels in the evaluation, which
is not enough for a~classif\/ication of the relationship between dif\/ferent models attempted in this
setting.

In fact, it is not clear in which sense~-- or if at all~-- spin-foam cosmology describes symmetric models.
It is true that inhomogeneous modes have not been truncated but remain present and may f\/luctuate,
potentially a~feature that would allow one to go beyond reduced models (see
also~\cite{SymmQFT,SymmStatesInt} in the canonical setting).
However, spin-foam cosmology at present lacks conditions that would ensure inhomogeneous modes to be
suf\/f\/iciently small for the models to be considered symmetric, not just in their f\/luctuations but even
in their expectation values.
The graphs used in spin-foam cosmology refer to f\/initely many degrees of freedom, often related
heuristically to the number of degrees of freedom of homogeneous minisuperspaces.
However, counting degrees of freedom is not enough to ensure that a~model is good.
One might simply def\/ine a~f\/inite-dimensional ``minisuperspace'' by picking some point $x_0$ in
space-time and considering only the metric components $g_{ab}(x_0)$ as degrees of freedom, a~model which
would be meaningless because of its dependence on the space-time gauge.
By using spatial embeddings, of graphs instead of points $x_0$, spin-foam cosmology is in danger of
producing models close to the one just sketched.
(Indeed, the status of covariance remains unclear in spin-foam cosmology as well as full spin foams.)

What is missing in this context is a~well-def\/ined analog of the map $\sigma$ for states used in loop
quantum cosmology.
Only such an object could tell whether the correct degrees of freedom have been captured.
Another question, related to the topics of this article, is how spin foams Abelianize.
The f\/inal equations often produced in this context resemble dif\/ference equations of Abelian loop
quantum cosmology, even though the starting point has SU(2) degrees of freedom.
No clear Abalianization step has been provided.
Finally, working with f\/ixed graphs embedded in space, spin-foam cosmology has not given rise to
ref\/inement models.

\subsection{Evaluation of models of quantum cosmology}

So far, we have discussed the construction of reduced and other models.
Their limited nature regarding the dynamics requires care also, and especially, when they are evaluated for
physi\-cal predictions.
In addition, there are caveats which apply to any construction in quantum gravity, and so to model systems
as well.
To guarantee that models, obtained in a~suf\/f\/iciently parameterized way to ensure their genericness, can
indeed be evaluated reliably, one must use evaluation methods or approximations that do not bring in hidden
assumptions about the ge\-ne\-ral form of ef\/fects.
Important questions such as the problem of time, deparameterization\footnote{Deparameterization has become
something of a~method of choice in quantum cosmology, and models in which physical Hilbert spaces are
derived in this way are often called ``complete quantizations''. However, as a~quantization of a~space-time
theory, such constructions can be considered complete only when one has shown that results do not depend on
the choice of internal time.
No such demonstration has been given in the models proposed so far.
See also the discussion in~\cite{ReducedKasner}.}, and potential signature change enter here.
Moreover, in spite of the ubiquitous use of ``ef\/fective equations'', in most cases they are based on
a~misinterpretation of the classical {\em limit} presented in~\cite[Section~4.3]{Bohr},
based on~\cite{SemiClass}, as a~semiclassical approximation.
Accordingly, important quantum corrections have been missed in many analyses: While some $\hbar$-terms are
kept in $\ell_{\rm P}$-related holonomy corrections, quantum back-reaction terms of the same order are
dropped.
For details on these important issues, we refer to the review~\cite{Stellenbosch}.

\section{Conclusions}

Much work remains to be done to establish reduced models of loop quantum gravity as well-def\/ined and
controlled approximations, and as reliable sources of detailed predictions in high-curvature regimes.
As discussed in Section~\ref{s:Reduction}, an analogous problem must be faced even classically in
relating non-symmetric geometries to symmetric ones: the averaging problem.
A complete understanding of quantum minisuperspace models as approximations of the full theory, even if
a~reduction mechanism is included, can be obtained only when the classical averaging problem is better
understood.
Lacking a~general solution, no approach to quantum gravity is yet able to produce a~complete derivation of
reduced models.

Nevertheless, with suf\/f\/icient care one can make progress and render it at least likely that all crucial
ef\/fects of the full theory are captured.
That non-Abelian ef\/fects should play some role in loop quantum cosmology and require caution has been
emphasized quite some time ago regarding specif\/ic properties of inverse-triad
corrections~\cite{BoundFull} as well as general properties of homogeneous models~\cite{DegFull}, but it has
not often been realized.
The present article of\/fers several new observations and constructions to this end: We have pointed out
that most considerations made so far in loop quantum cosmology suf\/fer from Abelian artefacts, related to
the use of function spaces on the Bohr compactif\/ication of the real line, following~\cite{Bohr}.
To correct this oversight, a~new quantization of homogeneous connections is developed in
Section~\ref{s:Mini}, which starts from non-Abelian models and takes into account the complete
structure of invariant connections.
The resulting Hilbert space representation, when restricted to Abelian variables, is related to
Bohr-quantized models by an isometric $*$-algebra morphism, but one that is not unitary or bijective.
We lose information when we map states to the Bohr--Hilbert space, corresponding to the edge-spin degeneracy
inherent in previous models.

The edge-spin degeneracy of holonomies is removed by the new quantization in non-Abelian and Abelian
models, giving a~better handle on lattice structures and the relation to the full theory.
We have strictly related basic operators in the full theory and in models, and showed how quantum ef\/fects
in composite operators can be captured by local quantizations.
The averaging required can lead to unexpected features~-- as seen in detail for f\/lux operators~-- which
one would not endeavor to implement in a~pure minisuperspace quantization without being prompted by the
relationship with the full theory.
Several implications have been demonstrated, especially regarding lattice ref\/inement and the form and
sizes of quantum-geometry corrections, most importantly those due to inverse triads.

We emphasize that our constructions started with the realization of def\/iciencies in current quantizations
based solely on Abelian models; seeing lattice ref\/inement or local features was not the main aim but
nevertheless resulted as an unavoidable consequence.
Section~\ref{s:limit} has provided cautionary remarks, detailing the current incomplete status of
the f\/ield and providing some guidelines for evaluations and the approach to physicality.

\subsection*{Acknowledgements}

The author is grateful to the anonymous referees for several helpful comments and suggestions.
This work was supported in part by NSF grants PHY-0748336 and PHY-1307408.

\pdfbookmark[1]{References}{ref}
\LastPageEnding

\end{document}